\documentclass[prd,aps,showpacs,notitlepage,superscriptaddress,nofootinbib]{revtex4-1}
\usepackage{graphicx}
\usepackage{url}
\usepackage{amsmath}
\usepackage{multirow}
\def\bea#1\eea{\begin{align}#1\end{align}}

\newcommand{\bef}{\begin{figure}[htb]\centering}
\newcommand{\eef}{\end{figure}}

\begin{document}
\title{Light and heavy flavor dijet production and dijet mass modification \\[.5ex] in heavy ion collisions}

\date{\today}

\author{Zhong-Bo Kang}
\email{zkang@physics.ucla.edu}
\affiliation{Department of Physics and Astronomy, University of California, Los Angeles, California 90095, USA}            
\affiliation{Mani L. Bhaumik Institute for Theoretical Physics, University of California, Los Angeles, California 90095, USA}
\affiliation{Theoretical Division, Los Alamos National Laboratory, Los Alamos, New Mexico 87545, USA}    

\author{Jared Reiten}
\email{jdreiten@physics.ucla.edu}
\affiliation{Department of Physics and Astronomy, University of California, Los Angeles, California 90095, USA}  
\affiliation{Mani L. Bhaumik Institute for Theoretical Physics, University of California, Los Angeles, California 90095, USA}

\author{Ivan Vitev}
\email{ivitev@lanl.gov}                   
\affiliation{Theoretical Division, Los Alamos National Laboratory, Los Alamos, New Mexico 87545, USA}       
                   
\author{Boram Yoon}
\email{boram@lanl.gov}
\affiliation{Theoretical Division, Los Alamos National Laboratory, Los Alamos, New Mexico 87545, USA}

\begin{abstract}
Back-to-back light and heavy flavor dijet measurements are promising experimental channels to accurately study the physics of jet production and propagation in a dense QCD medium. They can provide new insights into the path length, color charge, and mass dependence of quark and gluon energy loss in the quark-gluon plasma produced in reactions of ultra-relativistic nuclei. To this end, we perform a comprehensive study of both light and heavy flavor dijet production in heavy ion collisions. We propose the modification of dijet invariant mass distributions in such reactions as a novel observable that shows enhanced sensitivity to the quark-gluon plasma transport properties and heavy quark mass effects on in-medium parton showers. This is achieved through the combination of the jet quenching effects on the individual jets as opposed to their subtraction. The latter drives the subtle effects on more conventional observables, such as the dijet momentum imbalance shifts, which we also calculate here. Results are presented in Pb+Pb collisions at $\sqrt{s_{NN}}$ = 5.02~TeV for comparison to data at the Large Hadron Collider and in Au+Au collisions at $\sqrt{s_{NN}}$ = 200~GeV to guide the future sPHENIX program at the Relativistic Heavy Ion Collider.
\end{abstract}

\maketitle


\section{Introduction}
The high energy nuclear physics and particle physics communities are in the planning process for the upcoming proton and heavy ion runs at the Relativistic Heavy Ion Collider (RHIC) and the Large Hadron Collider (LHC). This is an opportune time to reflect on the success of recent runs and explore new opportunities. Important observables in high energy and heavy ion physics are related to hadronic jets~\cite{Salam:2009jx,Sapeta:2015gee}. For example, the dominant Higgs decay channel $H\rightarrow b\bar{b}$ was only recently observed~\cite{Aaboud:2017xsd,Sirunyan:2017elk} and involved $b$-jet reconstruction in the final state. There has also been a resurgence in the Quantum Chromodynamics (QCD) theory of jets with emphasis on their substructure, for a recent review of this physics in elementary p+p collisions see Ref.~\cite{Larkoski:2017jix}. In collisions of ultra-relativistic nuclei at RHIC and the LHC, the modification of the production cross sections and substructure of jets is more sensitive to the in-medium strong interaction dynamics in comparison to the leading hadron attenuation~\cite{Vitev:2008rz}. As such, jets are excellent diagnostics of the hot and dense medium, the quark-gluon plasma (QGP), that is created in heavy ion collisions (HIC). These jet quenching phenomena have been widely studied at both RHIC and the LHC, for a recent review of jet physics in HIC see Ref.~\cite{Connors:2017ptx}. 

Heavy flavor physics in reactions of ultra-relativistic nuclei is another incredibly active area of research~\cite{Andronic:2015wma} that predates the study of jets. For the purposes of this paper, we will restrict our discussion to open heavy flavor, where experimental measurements and related phenomenology have traditionally focused on $D$-meson and $B$-meson production. There is a great deal of interest in the use of heavy flavor to constrain the transport properties of the QGP~\cite{Rapp:2018qla,Cao:2018ews}. Still, the mechanisms of its in-medium modification are not yet fully understood. In light of this, heavy flavor jets have been proposed as a new tool to test the theory of heavy quark production, parton shower formation, and modification in nuclear matter. The first theoretical study of single inclusive $b$-jet production in HIC~\cite{Huang:2013vaa,Li:2018xuv} has found that the cross section receives a large contribution from prompt gluons, where heavy flavor emerges from gluon splitting only in the late stages of the parton shower evolution. Thus, the suppression of inclusive $b$-jets at high transverse momenta can be nearly as large as the quenching of light jets, as confirmed by the first CMS measurement~\cite{Chatrchyan:2013exa}. For the same reason, the connection between $b$-jet suppression and $b$-quark energy loss can be quite indirect. On the other hand, $B$-meson-tagged $b$-jets~\cite{Huang:2015mva} suppress such a contribution from gluon splitting, and are most effective in ensuring that the dominant fraction of recoiling jets originate from prompt $b$-quarks. Such a conclusion also applies to the back-to-back $b$-tagged dijet production, as we will show below. New measurements of heavy flavor jets and their substructure, which is particularly sensitive to mass effects on in-medium parton shower evolution~\cite{Li:2017wwc}, are expected from the upcoming LHC runs and from the future sPHENIX experiment at RHIC~\cite{Adare:2015kwa}.

Back-to-back jet pair (or dijet) production is among the most exciting channels used to probe QGP properties, where one typically focuses on the most energetic (``leading'') and second most energetic (``subleading'') jets. It is instructive to recall that the first definitive measurement of quenching effects on reconstructed jets came from the enhanced dijet asymmetry measurements at the LHC~\cite{Aad:2010bu,Chatrchyan:2011sx}. Further studies of this observable have been carried out not only at the LHC~\cite{Khachatryan:2015lha}, but also at RHIC~\cite{Adamczyk:2016fqm}. The origin of the additional imbalance to the dijet transverse momentum distribution in heavy ion collisions in comparison with the elementary p+p collisions has been attributed to the path length and color charge dependence of parton energy loss~\cite{Qin:2010mn,Young:2011qx,He:2011pd}. The interplay of Sudakov and in-medium collisional broadening on dijet acoplanarity has also been explored~\cite{Chen:2016cof}. More recently, efforts have been put forward to understand the dependence of the quenching on the type of parton that initiates the jet. The first measurement of the back-to-back $b$-jet momentum imbalance~\cite{Sirunyan:2018jju} has been performed at the LHC and modeled theoretically~\cite{Dai:2018mhw}.

The dijet asymmetry and momentum imbalance measure the {\em difference} of potentially large attenuation effects on the leading and subleading jets. Thus, those observables show a somewhat reduced sensitivity to the physics of jet quenching and the transport properties of the QGP. It has been pointed out early on that the asymmetry and momentum imbalance shifts in HIC may be influenced by background fluctuations~\cite{Cacciari:2011tm} and, more recently, by parton shower fluctuations on an event-by-event basis~\cite{Brewer:2018mpk}. To this end, we set out to find an observable where the effects that arise from the in-medium modification of parton showers {\em combine} rather than subtract, and lead to enhanced sensitivity to the interactions of jets in the QGP, as well as the mass dependence of parton energy loss. 

In the current work, we provide an extensive study of dijet production in heavy ion collisions at RHIC (or sPHENIX) kinematics and at LHC energies for both inclusive and $b$-tagged dijets. We compare the similarities and differences between those channels in A+A collisions to understand the flavor dependence of the quenching effects. Most importantly, we propose to use the dijet invariant mass modification as a novel probe of the QGP. An earlier study of dijet mass in proton-nucleus collisions showed negligible cold nuclear matter effects~\cite{He:2011sg}, suggesting that any significant modification of the dijet invariant mass distribution in A+A collisions arises from radiative and collisional energy loss processes in the QGP. At the same time, we include the studies for the more conventional observables such as two-dimensional nuclear modification factor $R_{AA}$ as a function of leading and subleading jet transverse momenta, and the imbalance $z_J$ distribution. We present theoretical predictions at $\sqrt{s_{NN}}$ = 200 GeV for future Au+Au collisions relevant to the sPHENIX kinematics at RHIC and at $\sqrt{s_{NN}}$ = 5.02 TeV for comparison to Pb+Pb data at the LHC. 

The rest of our paper is organized as follows. In Sec.~II, we present the evaluation of the differential cross sections for both inclusive and $b$-tagged dijet production in p+p collisions using the Pythia 8 event generator~\cite{Sjostrand:2007gs}. We also determine the flavor origin of the dijet production for the proper implementation of the energy loss effects. In Sec.~III, we first present the basic formalism used to generate dijet invariant mass distributions and imbalance distributions, starting from the double differential cross section in terms of the transverse momenta of leading and subleading jets. We then provide the information on how we implement the medium effects to obtain the modification of inclusive and $b$-tagged dijet production in dense QCD matter. In Sec.~IV, we present our phenomenological results for both RHIC and LHC kinematics. We give predictions for sPHENIX at RHIC, and provide detailed comparison with the most recent experimental measurements by the CMS collaboration at the LHC. We conclude our paper in Sec.~V. 

\section{Light and heavy flavor dijet production in p+p collisions}
In this section, we present the evaluation of the double differential cross sections for inclusive and $b$-tagged dijet production in p+p collisions using Pythia 8~\cite{Sjostrand:2007gs}, which is a widely-used high energy phenomenology event generator that describes the main properties of the event structure well. In our simulations, 8 million events are simulated for each of these two processes. We construct jets with the anti-$k_T$ jet clustering algorithm~\cite{Cacciari:2008gp}, where a $b$-jet is identified if there is at least one $b$-quark within the jet. 

\bef
\includegraphics[width=3.0in]{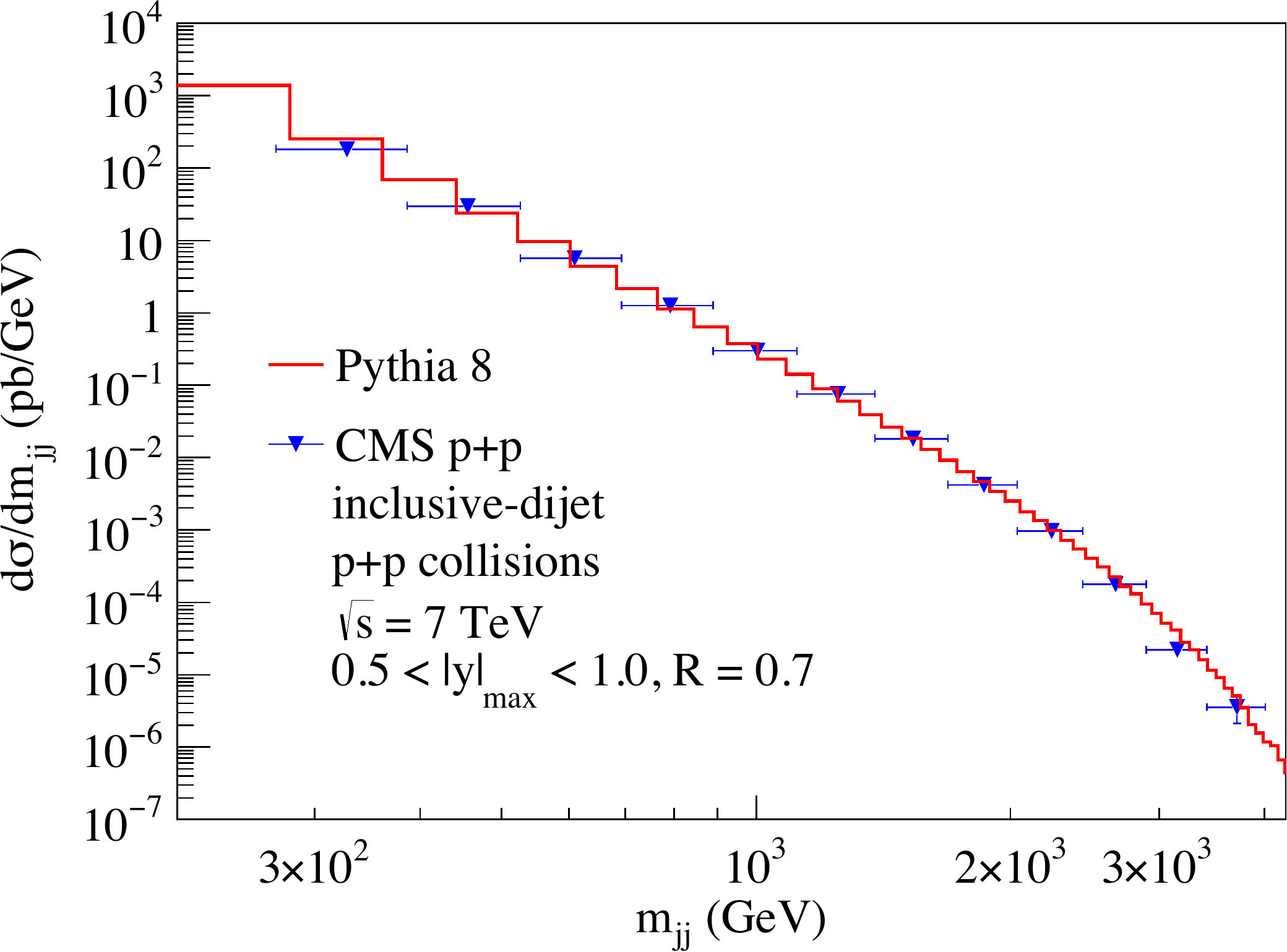}
\hskip 0.2in
\includegraphics[width=3.0in]{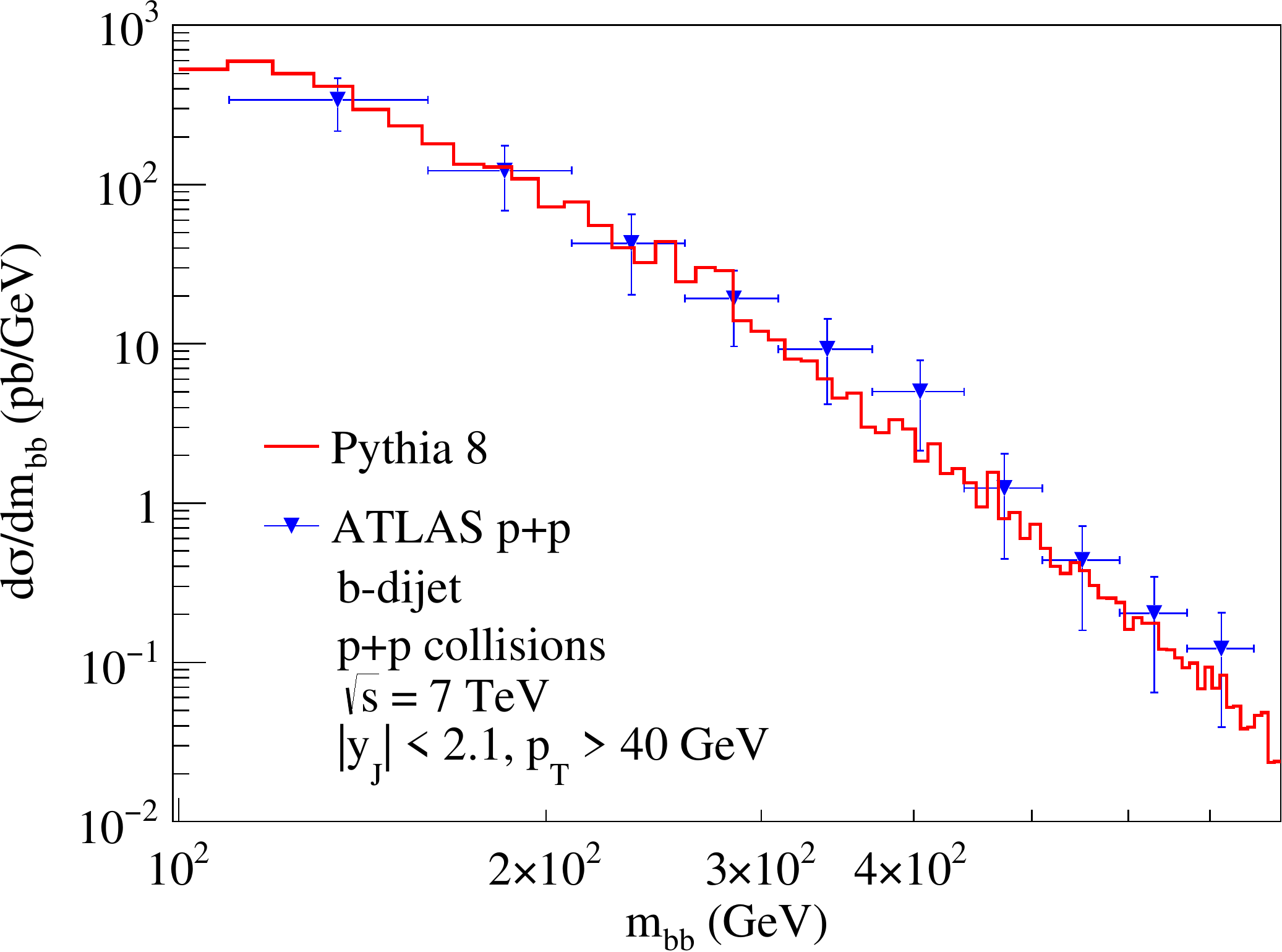}
\caption{Comparison of dijet mass distributions between Pythia 8 simulations and experimental measurements in p+p collisions at the LHC at $\sqrt{s} = 7$ TeV. The left is for inclusive dijets from the CMS collaboration~\cite{Chatrchyan:2012bja}, while the right is for $b$-tagged dijets from the ATLAS collaboration~\cite{ATLAS:2011ac}.}
\label{fig:mjj}
\eef

Both inclusive and $b$-tagged dijet production in p+p collisions have been measured at the LHC. To show the validation of Pythia 8 simulation against experimental measurements on inclusive dijet production in p+p collisions, we present dijet cross section as a function of dijet invariant mass in the left panel of Fig.~\ref{fig:mjj}, compared to experimental measurements by the CMS~\cite{Chatrchyan:2012bja} collaboration at center-of-mass energy $\sqrt{s} = 7$ TeV at the LHC. Here, the dijet invariant mass $m_{jj}$ is defined as
\bea
m_{jj}^2 = (p^{\rm L}+p^{\rm S})^2,
\eea
with $p^{\rm L}$ and $p^{\rm S}$ being the four-momenta for the leading and subleading jets, respectively. The jets are constructed with a jet radius $R=0.6$, along with the following rapidity cut
\bea
0.5 < y^* < 1.0,
\eea
where $y^* = |y^{\rm L} - y^{\rm S}|$ with $y^L$ ($y^S$) being the rapidity of leading and subleading jets. At the same time, we implement additional cuts on the transverse momentum and rapidity of individual jets, which are matched to those given in the experimental paper~\cite{Chatrchyan:2012bja}. The red histograms are the results from Pythia 8 simulations. As one can see, the Pythia 8 event generator describes the experimental dijet invariant mass data very well. This gives us confidence in extracting information on the parton flavors initiating the dijets in heavy ion collisions.

In the right panel of Fig.~\ref{fig:mjj}, we compare our Pythia 8 simulation for $b$-tagged dijet invariant mass distribution with the ATLAS measurement~\cite{ATLAS:2011ac} at $\sqrt{s} = 7$ TeV. The jet radius is $R = 0.4$ and the distance in rapidity and azimuthal angle between a $b$-quark and the $b$-jet, $\Delta R = \sqrt{(\Delta \eta)^2 + (\Delta \phi)^2}$, is required to be smaller than 0.3. Additionally, the transverse momentum of each $b$-quark is required to satisfy $p_T > 5$ GeV. All other event selection and kinematic cuts are implemented to match the experimental measurements. For details, see Ref.~\cite{ATLAS:2011ac}. Again, we obtain satisfactory agreement between our Pythia 8 simulation and the experimental data. 

\bef
\includegraphics[width=3.0in]{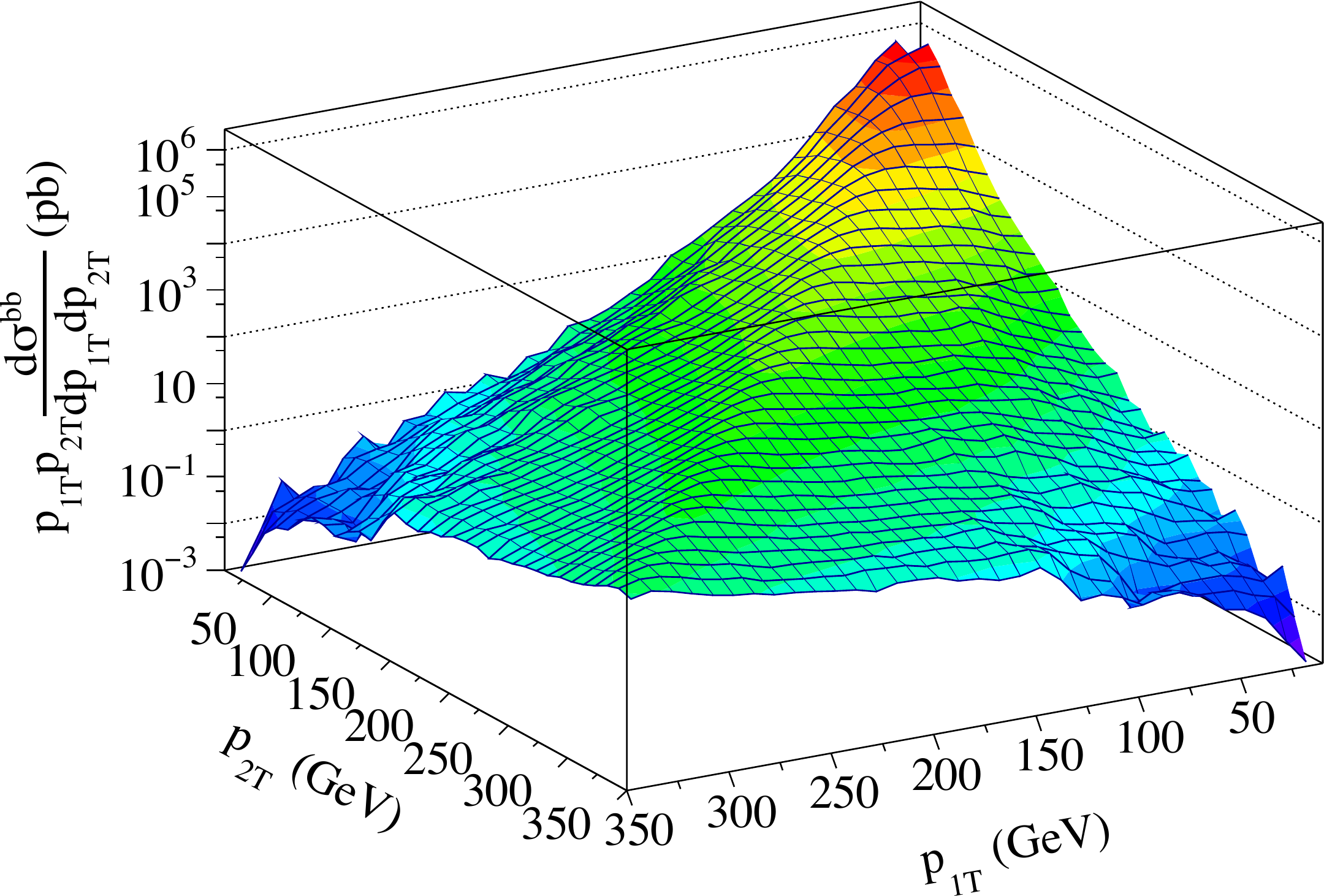}
\hskip 0.2in
\includegraphics[width=3.0in]{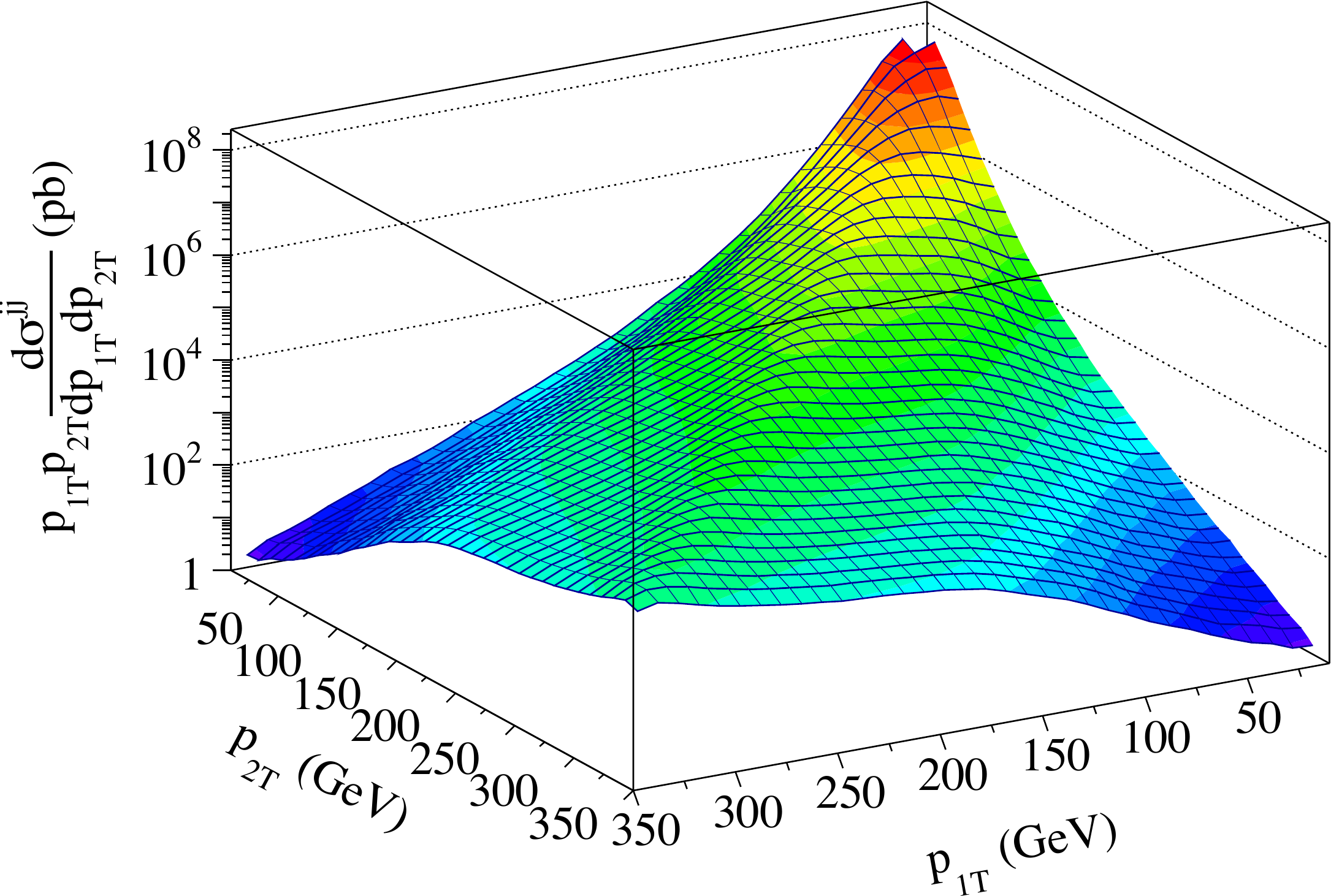}
\caption{Double differential cross sections weighted by transverse momenta for $b$-tagged (left) and inclusive (right) dijet production in p+p collisions at $\sqrt{s}=5.02$ TeV. Kinematic cuts are implemented in our simulations as in CMS measurements, see Ref.~\cite{Sirunyan:2018jju}. The roughness of the $b$-tagged dijet cross section relative to that for inclusive dijets is due to the inherently lower statistics.}
\label{fig:3D_LHC}
\eef

\bef
\includegraphics[width=3.0in]{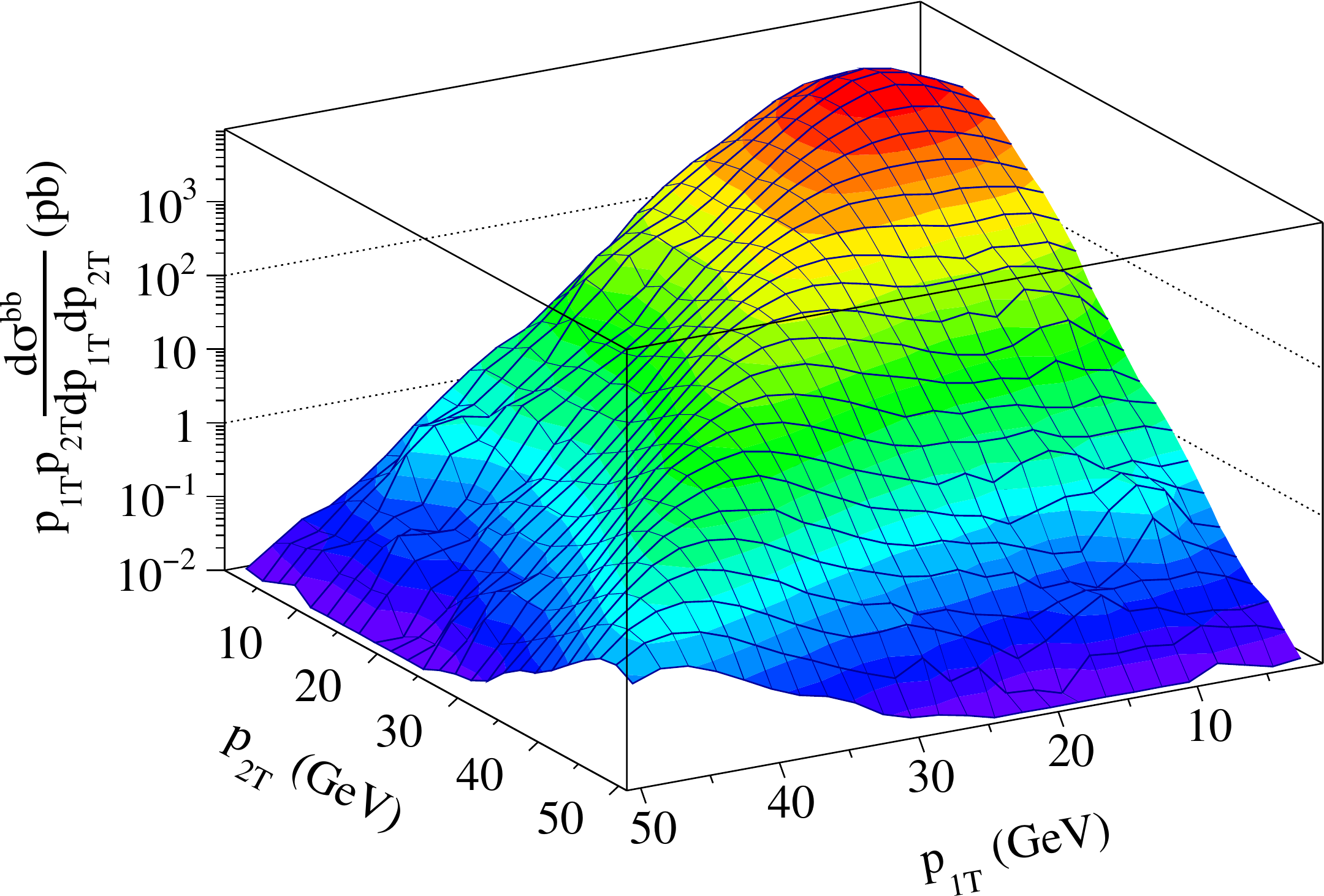}
\hskip 0.2in
\includegraphics[width=3.0in]{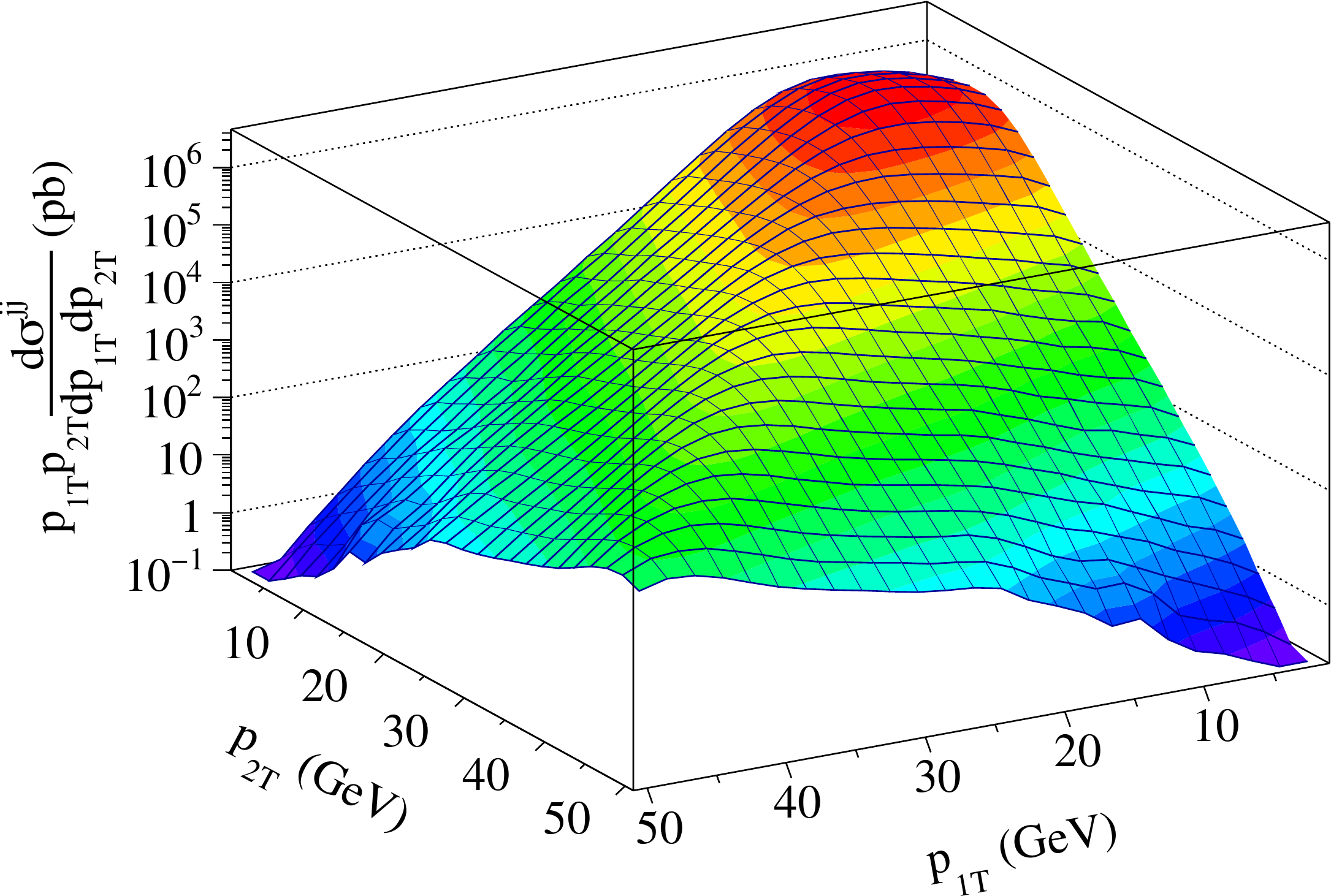}
\caption{Double differential cross sections weighted by transverse momenta for $b$-tagged (left) and inclusive (right) dijet production in p+p collisions at $\sqrt{s}=200$ GeV. Kinematic cuts implemented in our simulations are the same as those from the sPHENIX collaboration~\cite{sPHENIX}. Here again, the slight bumpiness of the $b$-tagged dijet cross section is due to its lower statistics relative to the inclusive case.}
\label{fig:3D_sPHENIX}
\eef

With the confidence that our comparisons to experimental measurements afford us, we now present the detailed baseline information for $b$-tagged and inclusive dijet production in p+p collisions, at $\sqrt{s} = 5.02$ TeV for the LHC and $\sqrt{s} = 200$ GeV for RHIC. These are the same center-of-mass energies (per nucleon pair) for the current heavy ion collisions at the LHC and for the planned sPHENIX experiment, respectively. 

In Fig.~\ref{fig:3D_LHC}, we show the three-dimensional (3D) plots of the cross section (weighted by the transverse momenta $p_{1T} p_{2T}$) at LHC energy $\sqrt{s} = 5.02$ TeV as a function of the transverse momenta of the two jets ($p_{1T}$ and $p_{2T}$) in the mid-rapidity region $|y|< 2$. The jets are reconstructed with a jet radius $R=0.4$. Here we label the dijet transverse momenta as $p_{1T}$ and $p_{2T}$ (instead of $p_T^{\rm L}$ and $p_T^{\rm S}$), because we do not distinguish which jet is leading or subleading in making the 3D plots. We will follow such a convention in the rest of the paper: when we need to specify leading and subleading jets, we label them as $p_T^{\rm L}$ and $p_T^{\rm S}$. Otherwise, we simply label them as $p_{1T}$ and $p_{2T}$. The left plot is for $b$-tagged dijet production, while the right is for inclusive dijets. Fig.~\ref{fig:3D_sPHENIX} is the same as Fig.~\ref{fig:3D_LHC}, but for sPHENIX energy $\sqrt{s} = 200$ GeV. The roughness of the $b$-tagged dijet cross section relative to that for inclusive dijets is due to the inherently lower statistics. As usual~\cite{He:2011pd}, the cross section reaches its maximum for $p_{1T}\approx p_{2T}$, and is broad and slowly varying as one goes away from this main diagonal. Such features will help us understand the behavior of nuclear modification in heavy ion collisions as we will see below. 

\bef
\includegraphics[width=3.0in]{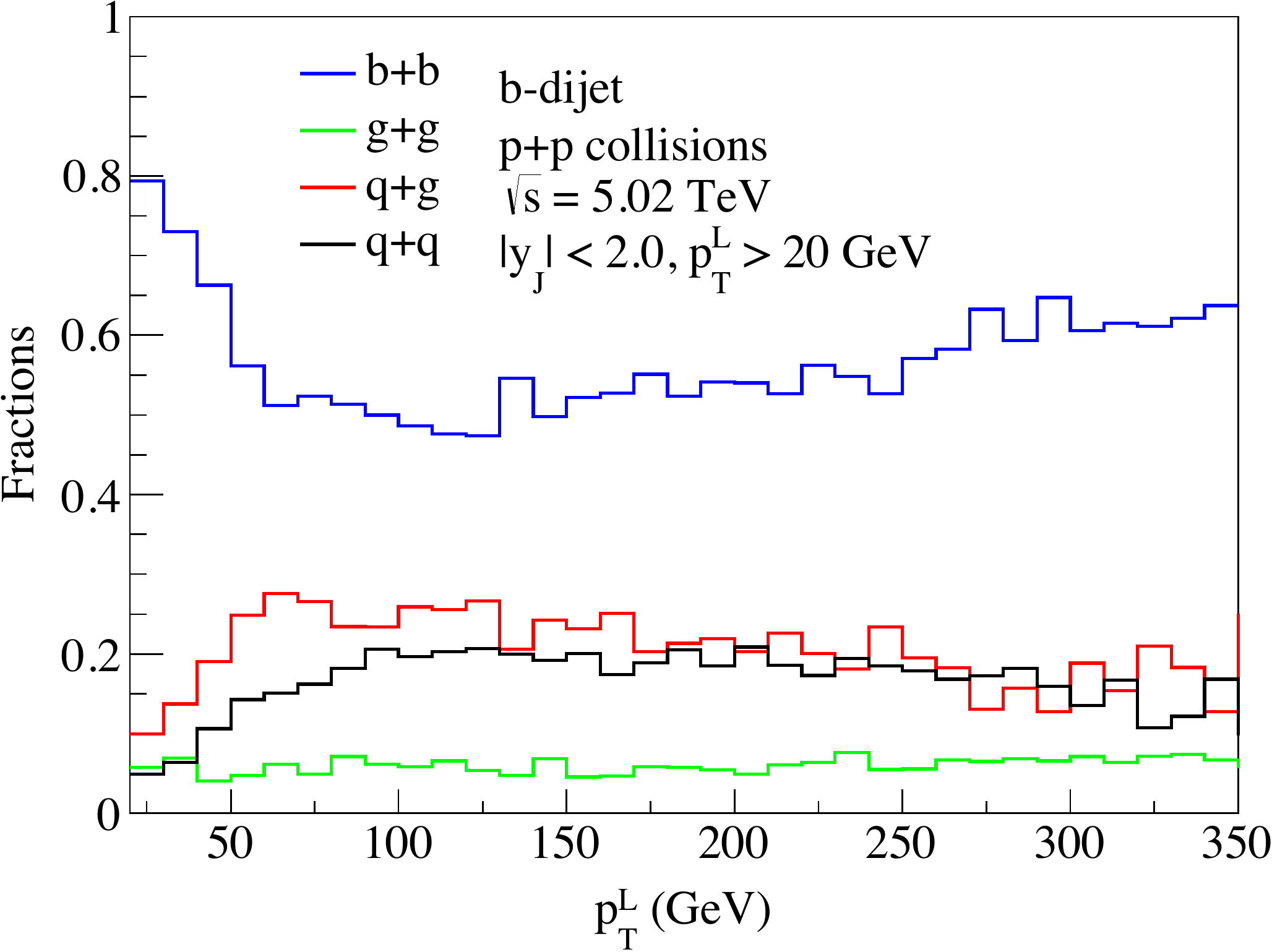}
\hskip 0.2in
\includegraphics[width=3.0in]{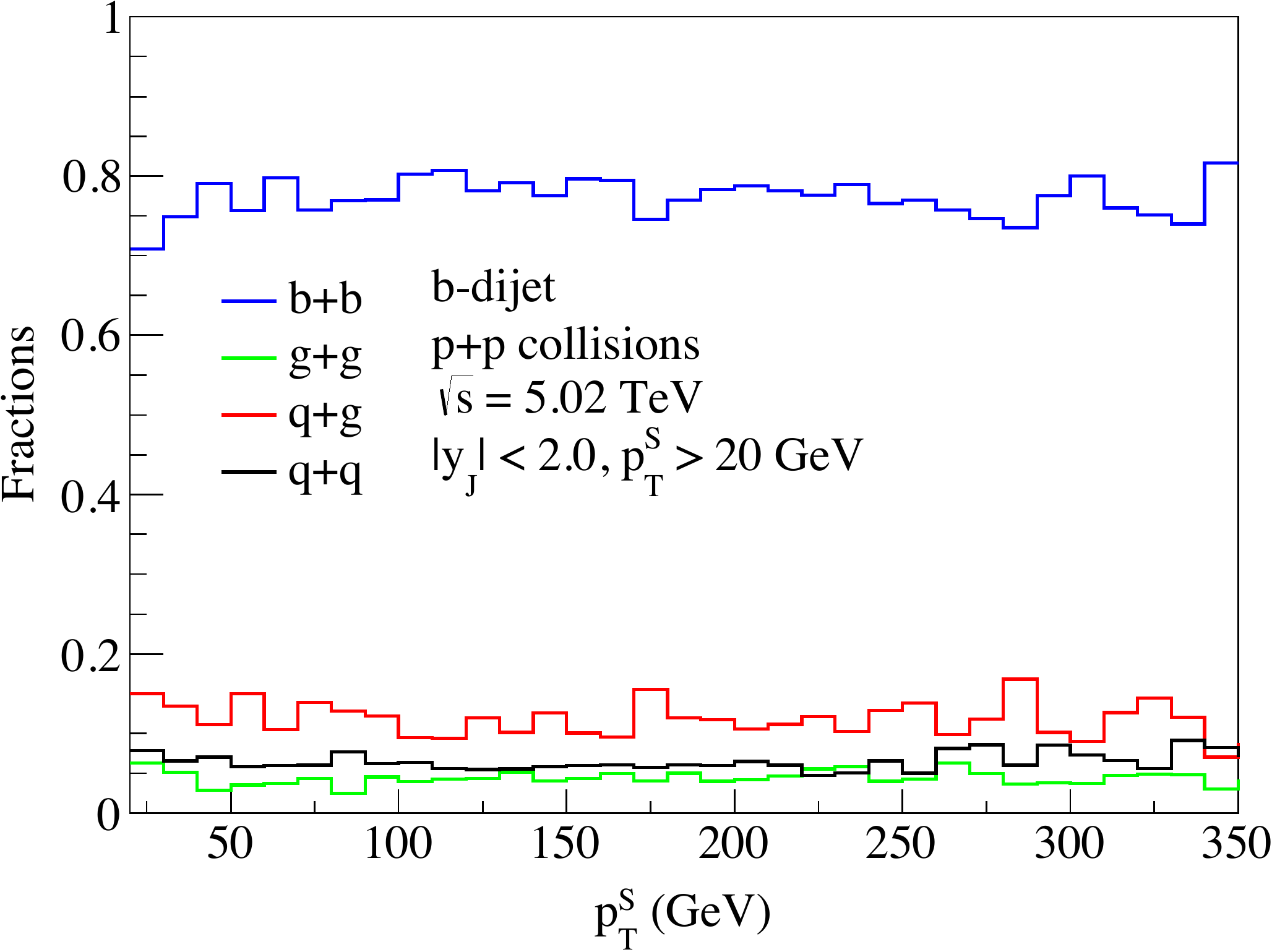}
\caption{The fractional contributions of different subprocesses to the $b$-dijet production cross sections vs. leading jet $p_T^{\rm L}$ (left) and subleading jet $p_T^{\rm S}$ (right) in p+p collisions at $\sqrt{s}=5.02$ TeV. Kinematic cuts are implemented in our simulations as in CMS measurements~\cite{Sirunyan:2018jju}.}
\label{fig:fracs_b}
\eef

\bef
\includegraphics[width=3.0in]{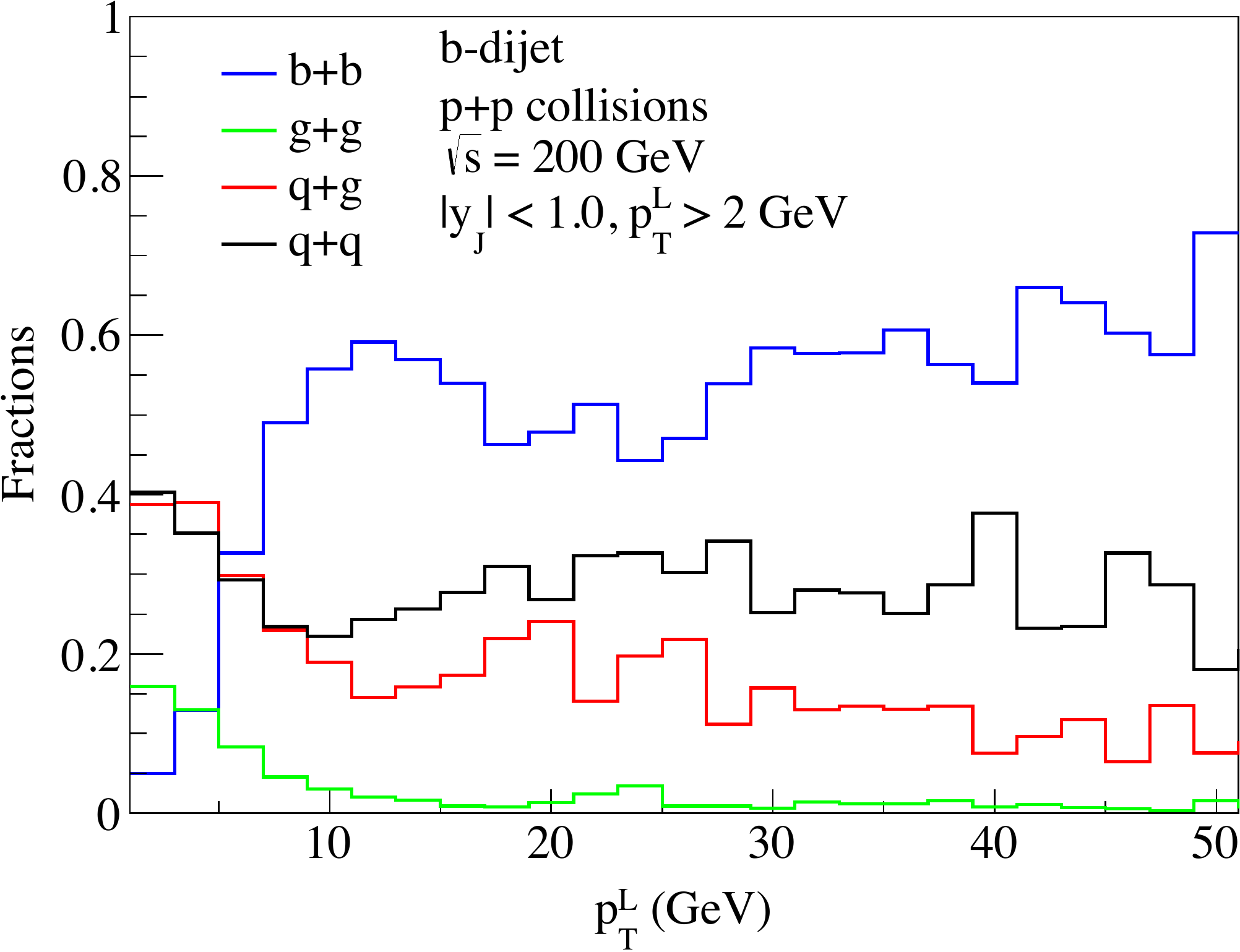}
\hskip 0.2in
\includegraphics[width=3.0in]{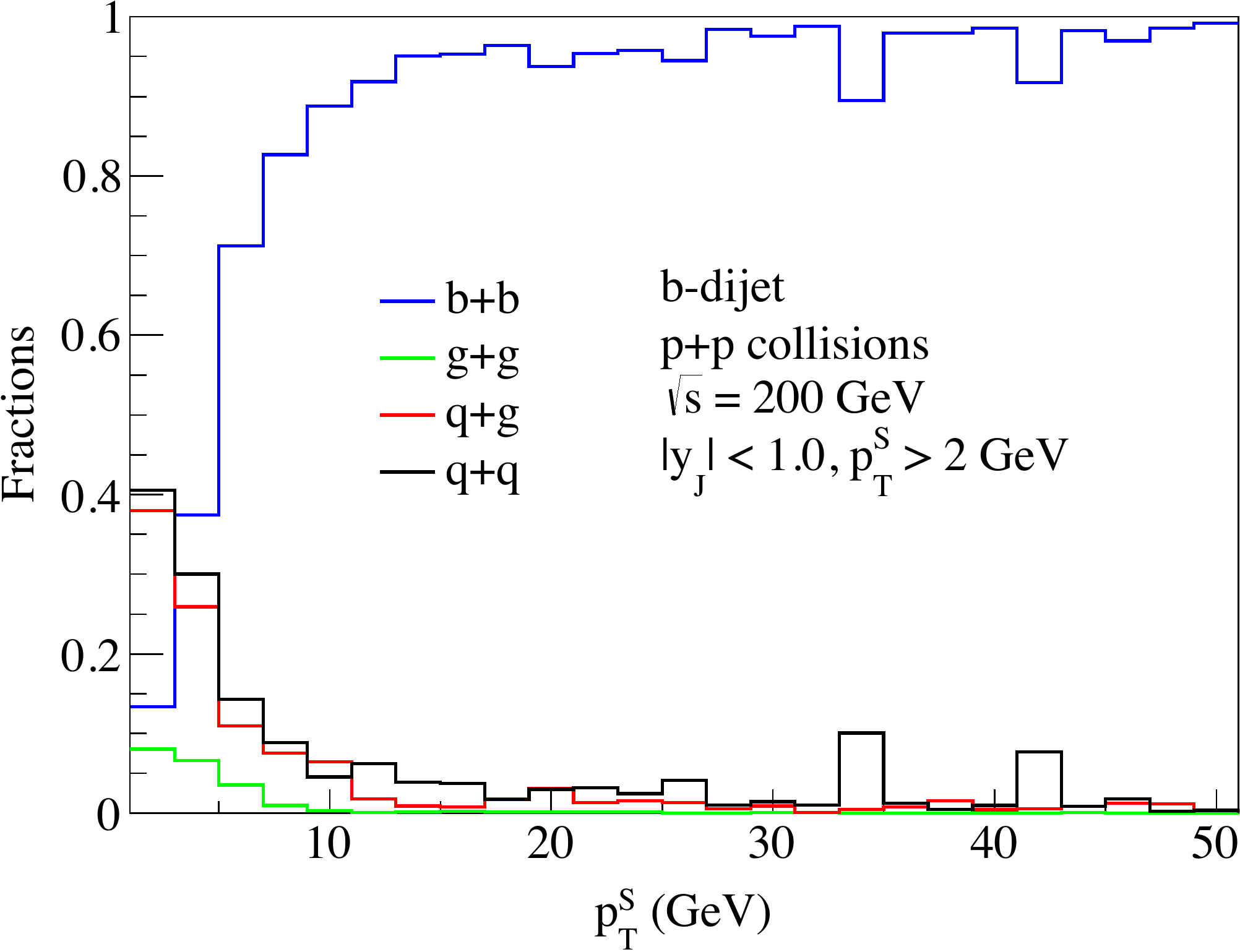}
\caption{The fractional contributions of different subprocesses to the $b$-dijet production cross sections vs. leading $p_T$ (left) and subleading $p_T$ (right) in p+p collisions at $\sqrt{s}=200$ GeV. Kinematic cuts implemented in our simulations are the same as those from the sPHENIX collaboration~\cite{sPHENIX}.}
\label{fig:fracs_b_phenix}
\eef

Let us now turn to the flavor origin of the dijets, which will be of central importance for our simulations of medium effects in heavy ion collisions, presented in the next section. The detailed kinematic constraints are shown in each plot. Pythia 8 utilizes leading order (LO) perturbative QCD matrix elements combined with parton showers. For $b$-tagged dijet production, there are 7 channels in our simulations: $g+g\to b+\bar b,~ q+\bar q \to b+\bar b,~ g+g \to g+g, ~ q+\bar q \to g+g,~ q+g\to q+g,~ g+g\to q+\bar q, q+q\to q+q$. We classify these 7 channels to 4 subprocesses according to the flavor information of the final state partons in LO matrix elements: (1) $g+g\to b+\bar b$, $q+\bar q \to b+\bar b$; (2) $g+g \to g+g$, $q+\bar q \to g+g$; (3) $q+g\to q+g$; (4) $g+g\to q+\bar q$, $q+q\to q+q$. We show in Figs.~\ref{fig:fracs_b} and \ref{fig:fracs_b_phenix} the fractions of these 4 subprocesses as functions of leading (trigger) jet $p_T^{\rm L}$ and subleading (associate) jet $p_T^{\rm S}$ at $\sqrt{s} = 5.02$~TeV and $\sqrt{s} = 200$ GeV, respectively.

The blue line labeled as $b+\bar b$ denotes the contributions from category (1), with $b+\bar b$ in the final state. In this case, both $b$-tagged jets are initiated by either a $b$-quark or a $\bar b$-quark. In heavy ion collisions, the medium modification of such $b$-jets has a direct connection to the physical heavy quark energy loss (mass $m_b$). The green curve labeled as $g+g$ includes the contributions from category (2), with $g+g$ in the final state. In this case, both $b$-tagged jets are initiated by prompt gluons through $g\to b+\bar b$ splitting in the showering process. Thus, the medium modification of these $b$-jets would resemble that of a massive gluon of effective mass $2m_b$. Similarly, the red curve denotes the process from category (3). Thus, one $b$-jet is initiated by a gluon $g$ like above. On the other hand, the other $b$-jet is initiated by a light quark $q$, for which the medium modification would resemble that of a massive quark. Finally, the black curve denotes the processes in category (4). In this case, both of the $b$-tagged jets are initiated by light quarks~$q$. As we can see, for a wide kinematic coverage, the subprocesses with $b+\bar b$ in the final state provide the dominant contributions ($\gtrsim 50 \%$) to $b$-tagged dijet production in p+p collisions at the LHC at $\sqrt{s} = 5.02$ TeV. On the other hand, the $b + \bar{b}$ channel dominates $b$-tagged dijet production across the $p_T$ range above 10 GeV, which is the relevant range for the sPHENIX experiment. This indicates that $b$-tagged dijet production provides an excellent opportunity to study the effects of heavy quark energy loss in heavy ion collisions.

\bef
\includegraphics[width=3.0in]{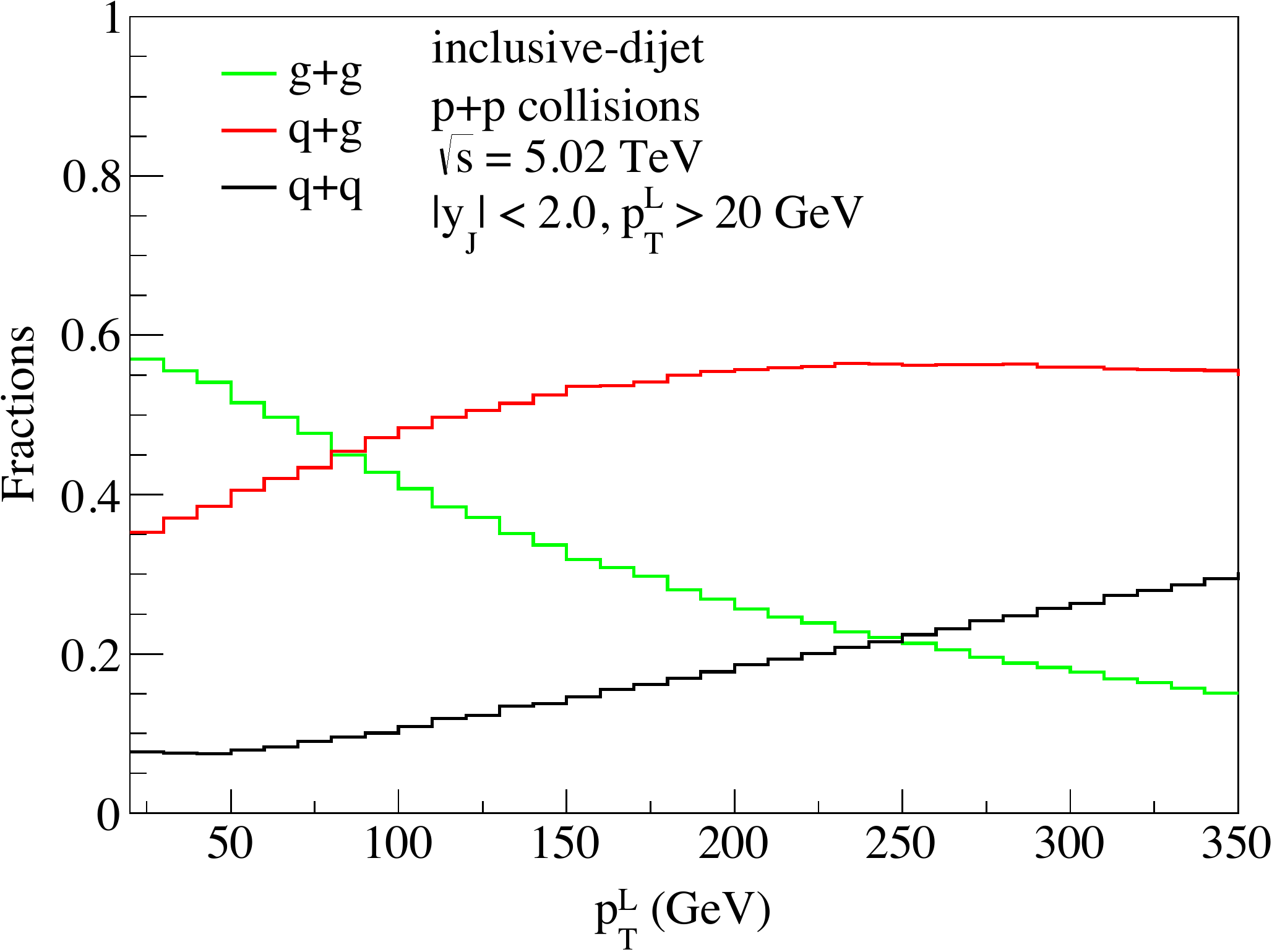}
\hskip 0.2in
\includegraphics[width=3.0in]{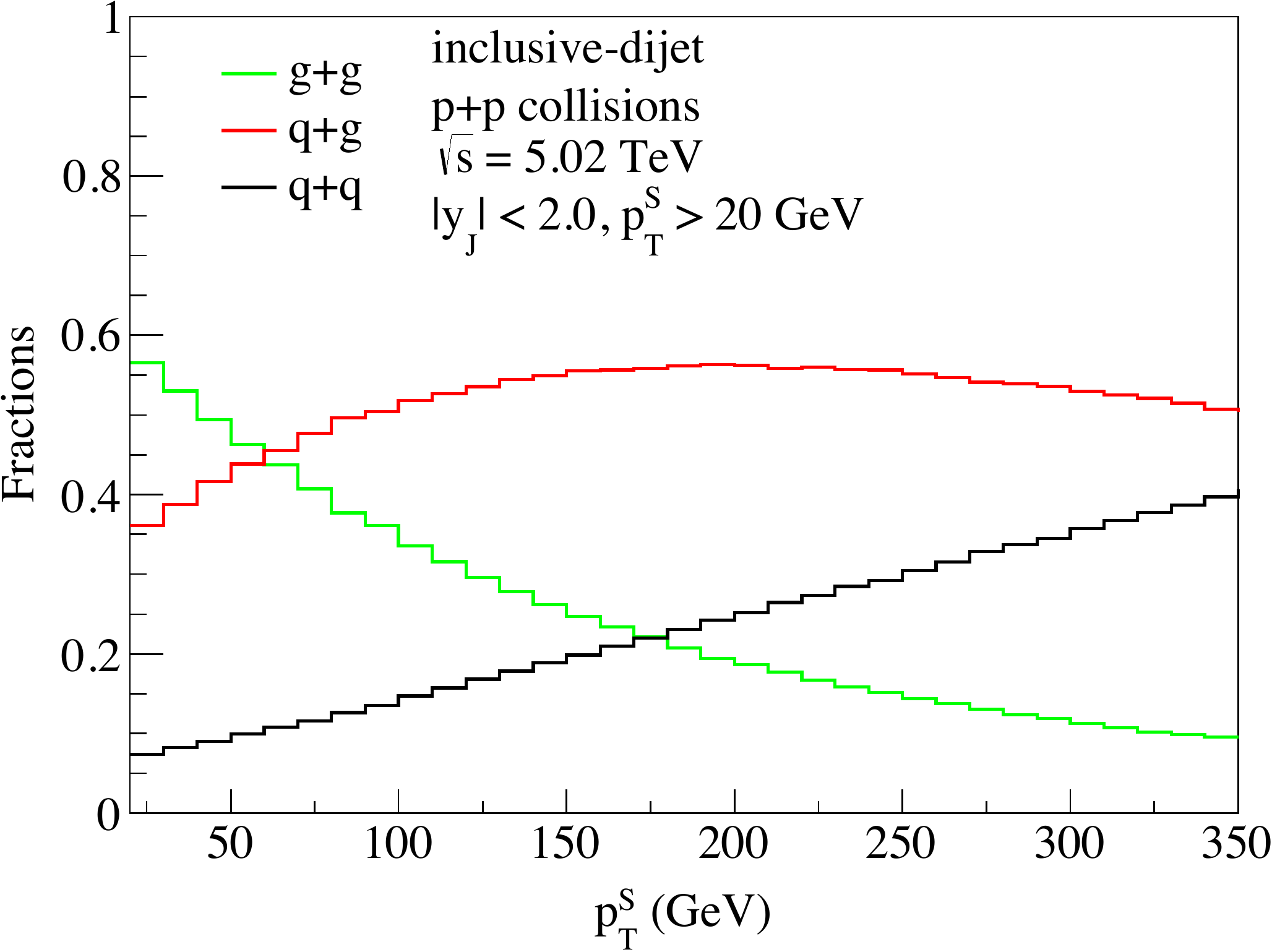}
\caption{The fractional contributions of different subprocesses to the inclusive dijet production cross sections vs. leading $p_T$ (left) and subleading $p_T$ (right) in p+p collisions at $\sqrt{s}=5.02$ TeV. Kinematic cuts are implemented in our simulations as in CMS measurements~\cite{Sirunyan:2018jju}.}
\label{fig:fracs_i_LHC}
\eef

\begin{figure}[hbt]
\includegraphics[width=3.0in]{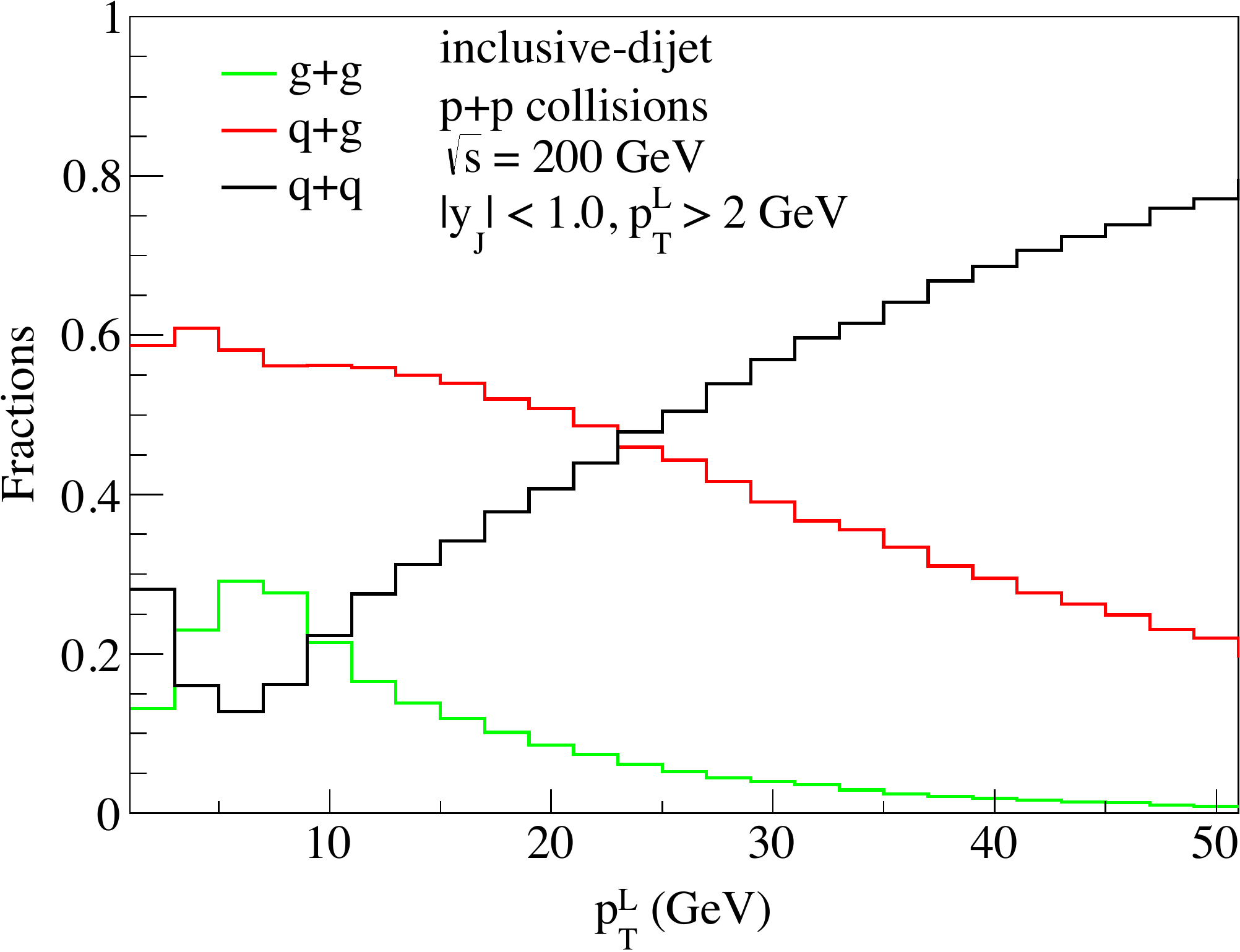}
\hskip 0.2in
\includegraphics[width=3.0in]{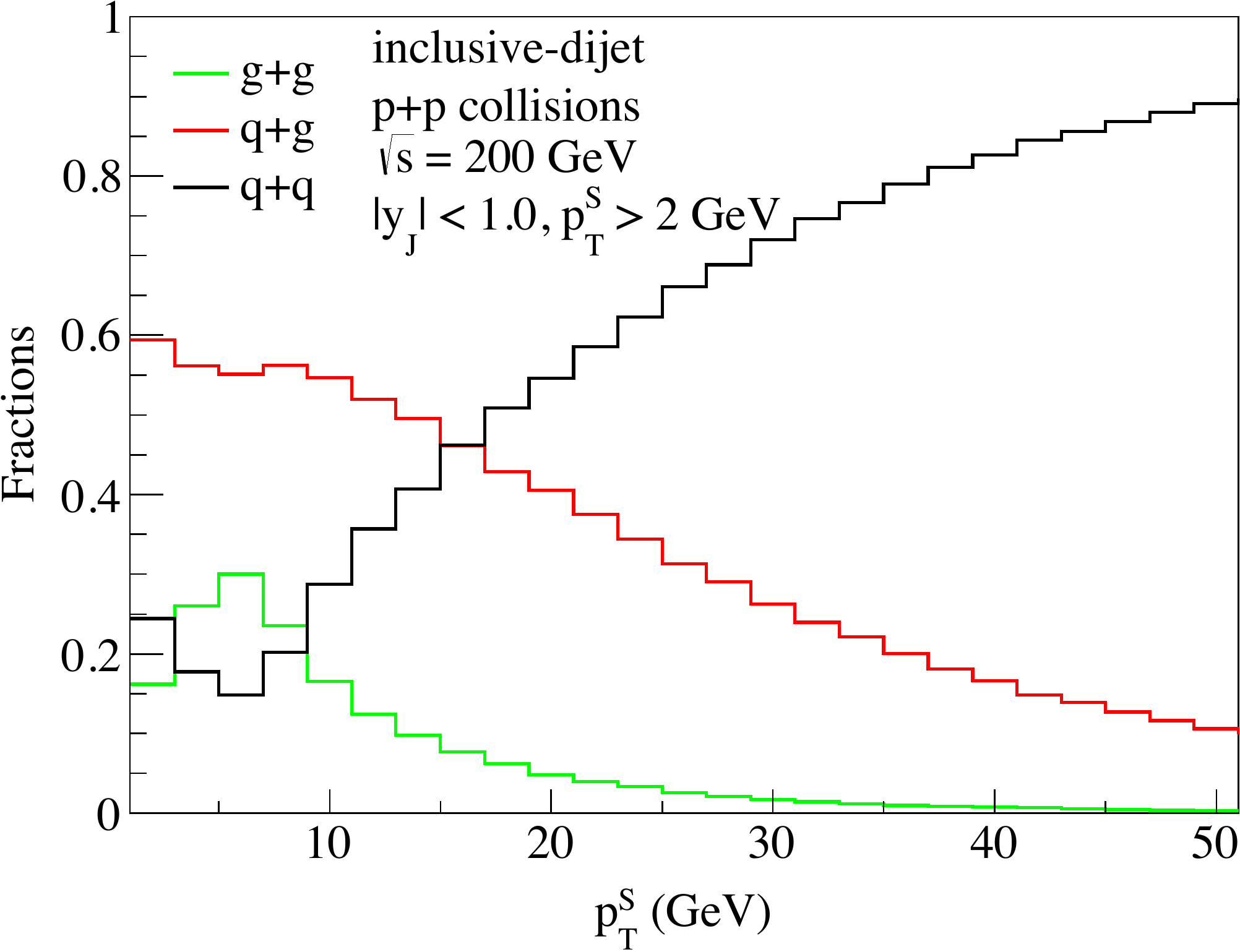}
\caption{The fractional contributions of different subprocesses to the inclusive dijet production cross sections vs. leading $p_T$ (left) and subleading $p_T$ (right) in p+p collisions at $\sqrt{s}=200$ GeV. Kinematic cuts implemented in our simulations are the same as those from the sPHENIX collaboration~\cite{sPHENIX}.}
\label{fig:fracs_i_sPHENIX}
\eef

On the other hand, for inclusive dijet production, the usual 5 partonic processes will be reclassified into three subprocesses through their final state parton contents: (1) $g+g \to g+g, ~ q+\bar q \to g+g$; (2) $q+g\to q+g$; (3) $g+g\to q+\bar q, q+q\to q+q$. In category (1), both jets are initiated by gluons, while for category (3), both jets are initiated by quarks. For category (2), the dijets are initiated by a light quark $q$ and a gluon $g$, respectively. One can clearly see in Fig.~\ref{fig:fracs_i_LHC} that at LHC energy $\sqrt{s} = 5.02$ TeV, for a large kinematic region, the process from category (2) is the dominant channel for inclusive dijet production. In other words, inclusive dijets at LHC kinematics are mostly initiated by a quark $q$ on one side and a gluon $g$ on the other end of the azimuthal plane. In addition, we plot such fractions in Fig.~\ref{fig:fracs_i_sPHENIX} at sPHENIX energy $\sqrt{s} = 200$ GeV as a function of leading jet transverse momentum $p_T^{\rm L}$ (left panel) and of the subleading jet transverse momentum $p_T^{\rm S}$ (right panel), respectively. We find that at relatively lower jet transverse momenta ($\lesssim 20$ GeV), the inclusive dijet cross section is dominated by category (2) with $q+g$ in the final state. At the higher jet transverse momenta, the cross section is dominated by category (3) with $q+q$ in the final state. This is expected since as the jet transverse momenta increase, the parton momentum fractions $x$ in the protons reach the region $x\sim 1$, where valence quarks dominate.

\section{Light and heavy flavor dijet production in hot QCD matter}
In this section, we provide the main formula and basic information on how we implement parton energy loss for both inclusive and $b$-tagged dijet production in heavy ion collisions. 

\subsection{Dijet production: main formula}
\label{sec:main formula}
Our starting point for both p+p and A+A collisions is the double differential cross section, ${d\sigma}/{dp_{1T}dp_{2T}}$, in two-dimensional transverse momentum bins $(p_{1T}, p_{2T})$ of the leading and subleading jets. With such a double differential cross section at hand, one can compute the dijet invariant mass distribution, as well as the so-called imbalance distribution as follows. 

The dijet invariant mass $m_{12}^2 = (p_1 + p_2)^2$ can be written in terms of the jet transverse momentum and rapidity as follows
\bea
m_{12}^2 = m_1^2 + m_2^2 + 2\left[m_{1T}m_{2T}\mathrm{cosh}(\Delta \eta) - p_{1T}p_{2T}\mathrm{cos}(\Delta \phi)\right],
\eea
where $m_1^2 = p_1^2$ and $m_{1T} = \sqrt{m_1^2 + p_{1T}^2}$ are the invariant mass squared and the transverse mass for one of the jets, likewise we have $m_2$ and $m_{2T}$ for the other jet. At the same time, we have the difference in the rapidities and the azimuthal angles as 
\bea
\Delta \eta = \eta_1 - \eta_2,
\qquad
\Delta \phi = \phi_1 - \phi_2,
\eea
where $\eta_{1,2}$ and $\phi_{1,2}$ are the rapidities and azimuthal angles for the jets. In the relevant kinematic regimes where the transverse momentum is much larger than the jet mass, $p_{T} \gg m$, we approximate $m_{T}\approx p_T$ and obtain
\bea
m_{12}^2 \approx m_1^2 + m_2^2 + 2p_{1T}p_{2T}\left[\mathrm{cosh(\Delta \eta)} - \mathrm{cos}(\Delta \phi)\right].
\label{eq:massform}
\eea
In the actual Pythia 8 simulations for dijet production, we generate the averaged $\langle m_1^2\rangle$, $\langle m_2^2\rangle$, and $\langle \mathrm{cosh(\Delta \eta)} - \mathrm{cos}(\Delta \phi)\rangle$ for each $(p_{1T}, p_{2T})$ bin. With this information, we compute the dijet invariant mass distribution through the double differential dijet momentum distribution via the following formula
\bea
\frac{d\sigma}{dm_{12}} = \int dp_{1T} dp_{2T} \frac{d\sigma}{dp_{1T}dp_{2T}} \delta\left(m_{12} - \sqrt{\langle m_1^2\rangle + \langle m_2^2\rangle + 2p_{1T}p_{2T}\langle\mathrm{cosh(\Delta \eta)} - \mathrm{cos}(\Delta \phi)\rangle} \right),
\label{eq:mass}
\eea
where the transverse momenta $p_{1T}$ and $p_{2T}$ are integrated over the desired experimental cuts. 

\begin{figure}[hbt]
\includegraphics[width=3.0in]{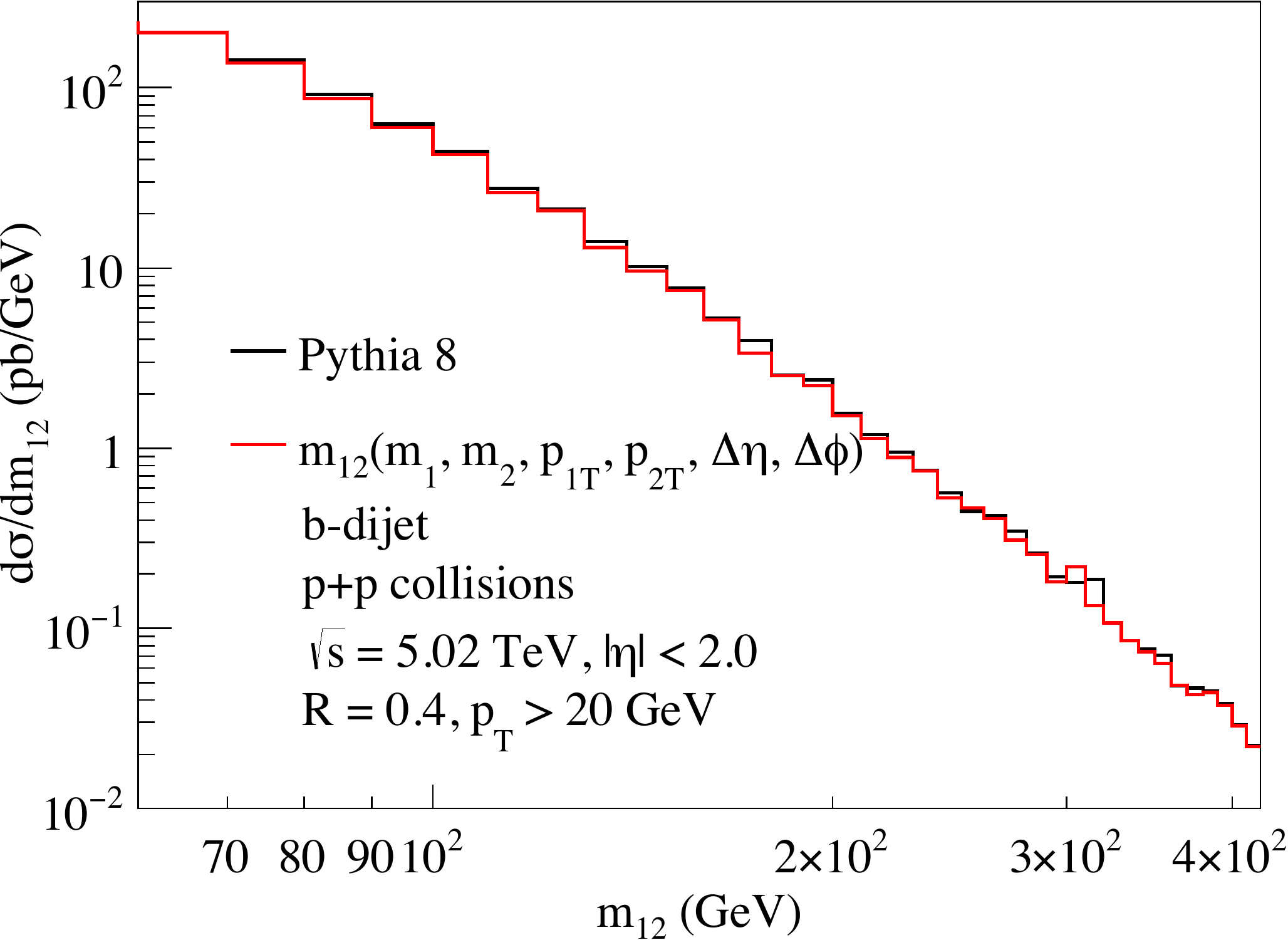}
\hskip 0.2in
\includegraphics[width=3.0in]{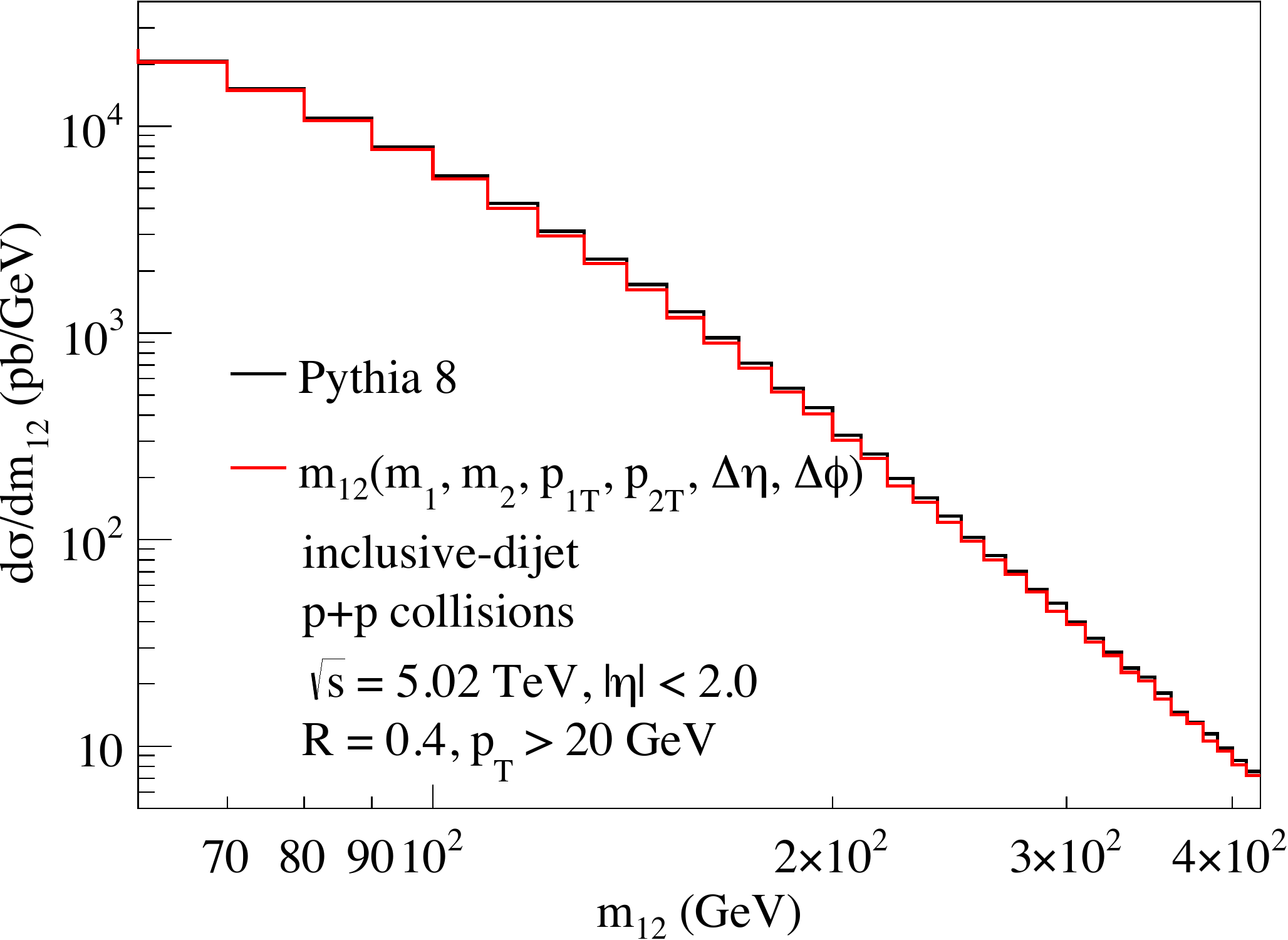}
\vskip 0.2in
\includegraphics[width=3.0in]{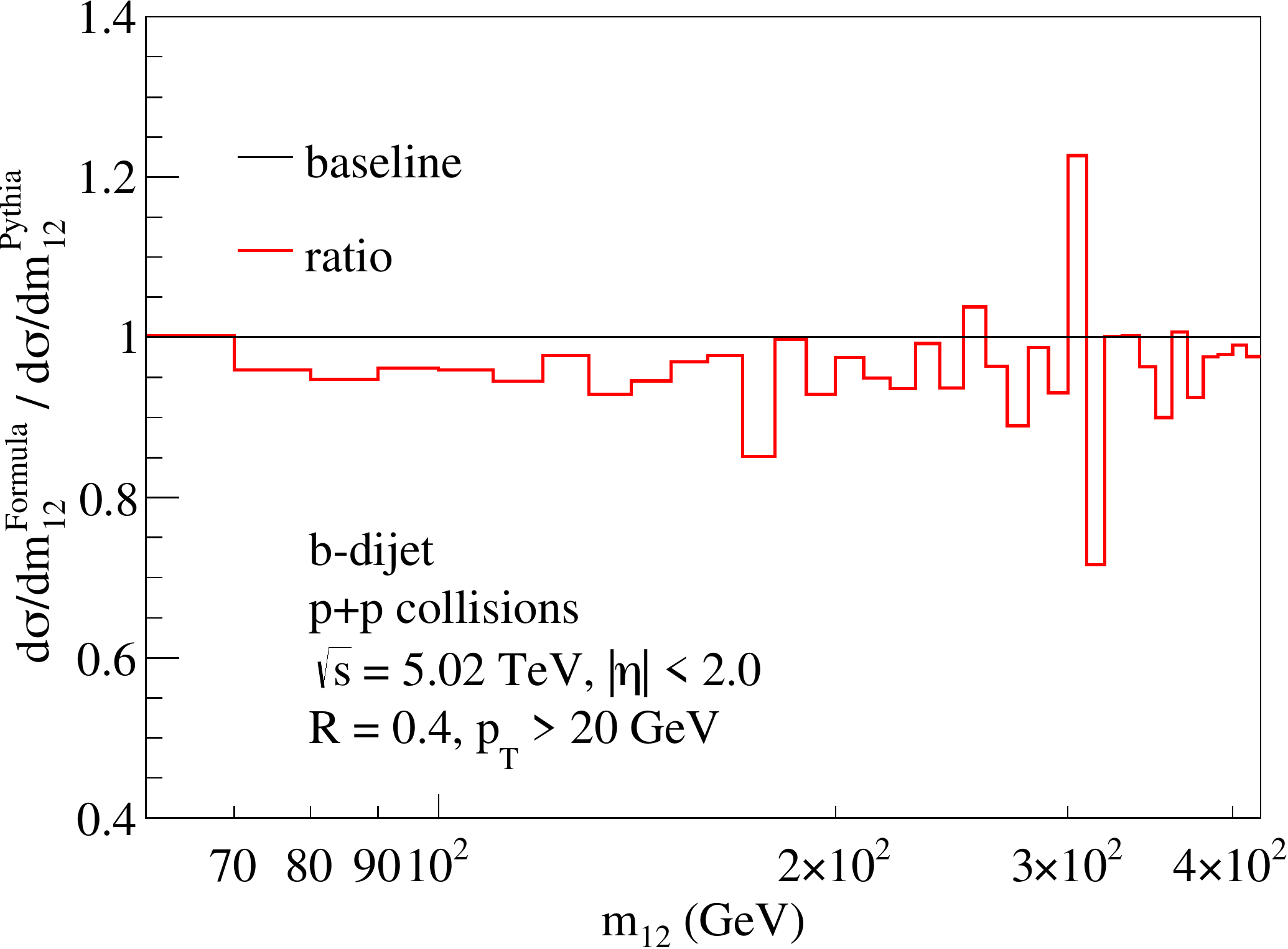}
\hskip 0.2in
\includegraphics[width=3.0in]{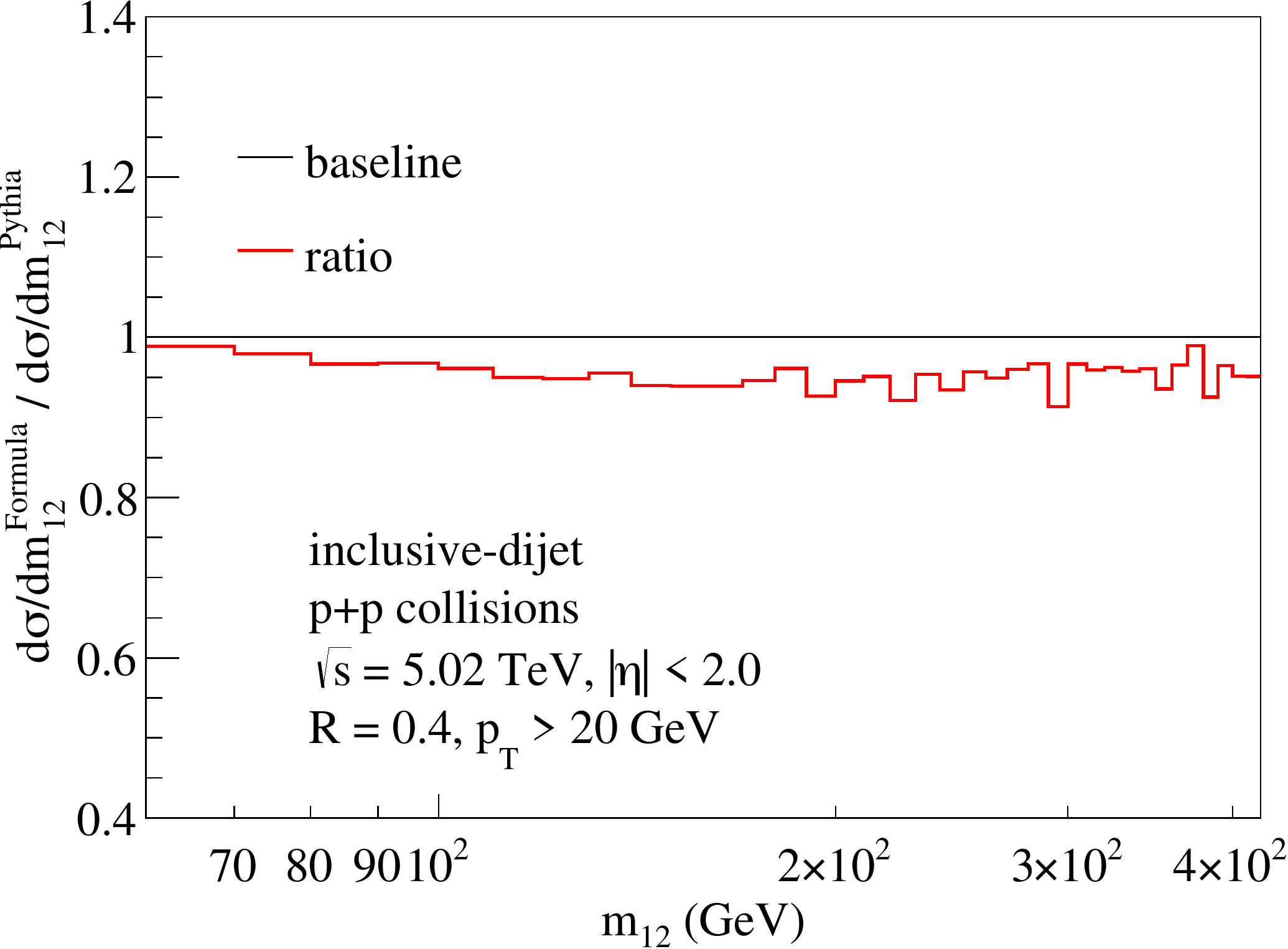}
\caption{Mass distributions (top) and their ratios (bottom) for $b$-tagged (left) and inclusive (right) dijet production in p+p collisions at $\sqrt{s}=5.02$ TeV. Kinematic cuts are implemented in our simulations as in CMS measurements~\cite{Sirunyan:2018jju}. The upper panels display black histograms representing $d\sigma/dm_{12}$ simulated {\it directly} from Pythia 8, while the red histograms are $d\sigma/dm_{12}$ computed using Eq.~\eqref{eq:mass} and Pythia 8-simulated~$d\sigma/dp_{1T}dp_{2T}$. In the lower panels the Pythia calculation is given by the black lines  and the red histograms represent the ratio  of the approximate formula to the baseline.}
\label{fig:mass_check_LHC}
\eef

\begin{figure}[hbt]
\includegraphics[width=3.0in]{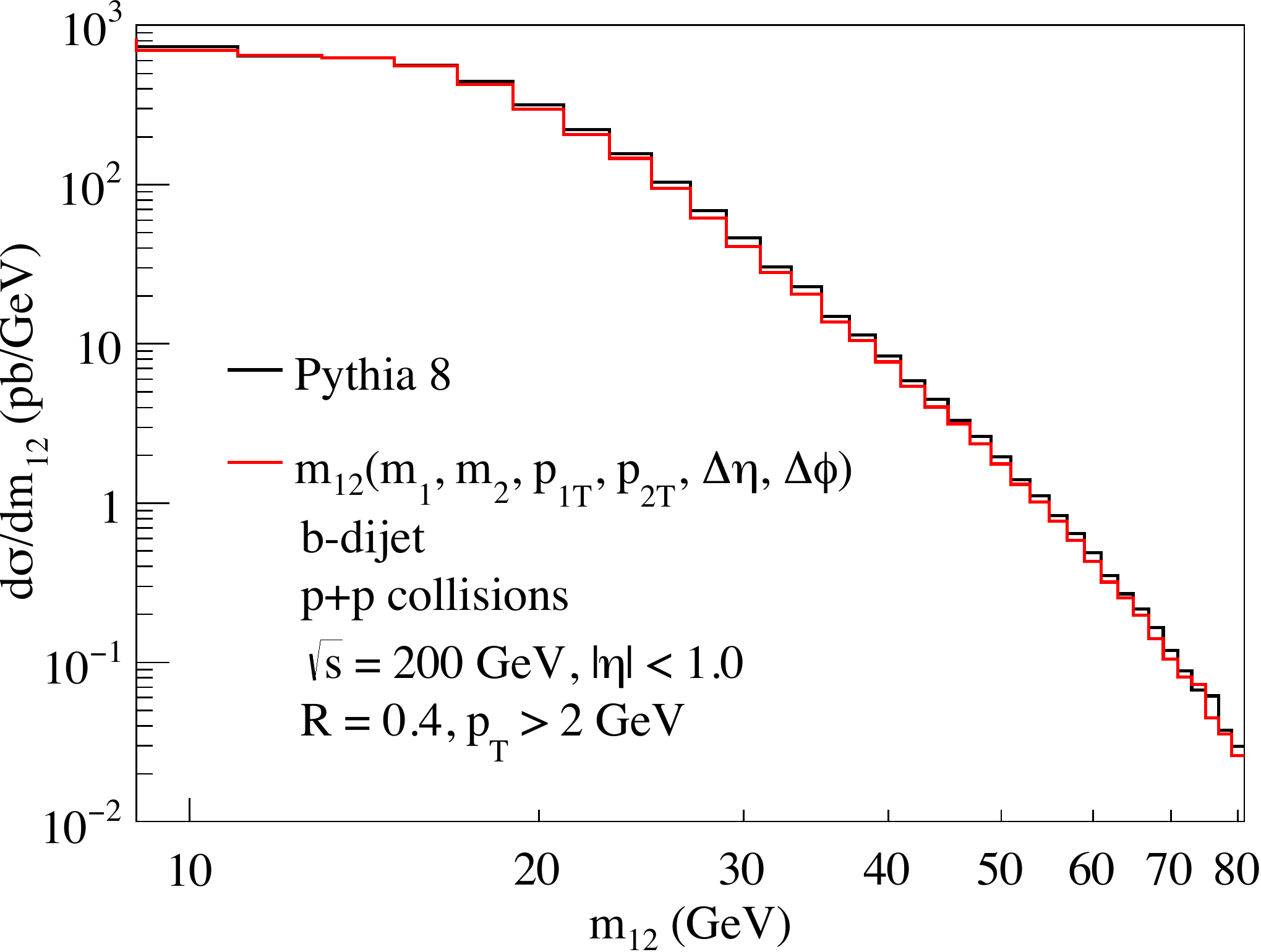}
\hskip 0.2in
\includegraphics[width=3.0in]{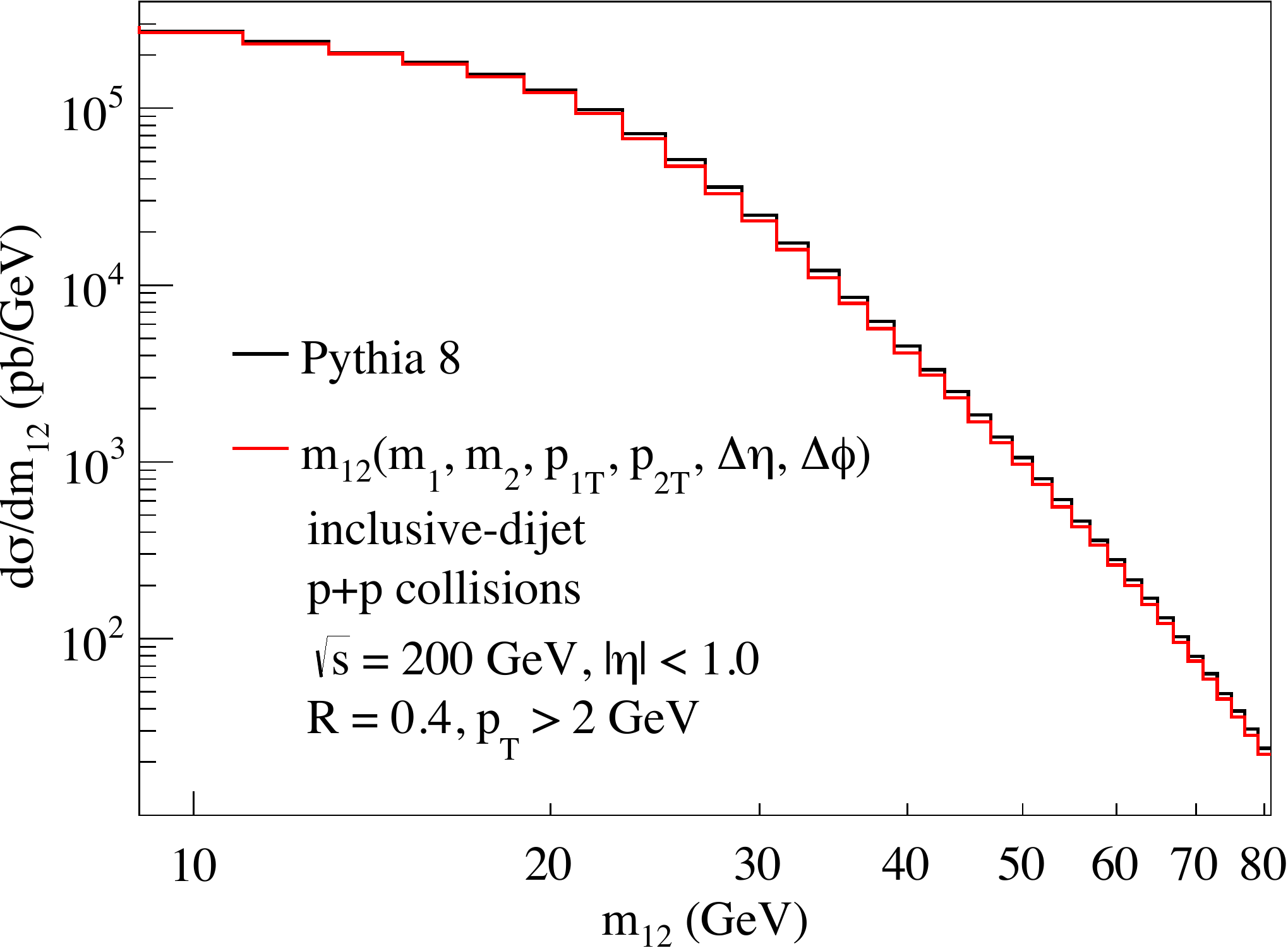}
\vskip 0.2in
\includegraphics[width=3.0in]{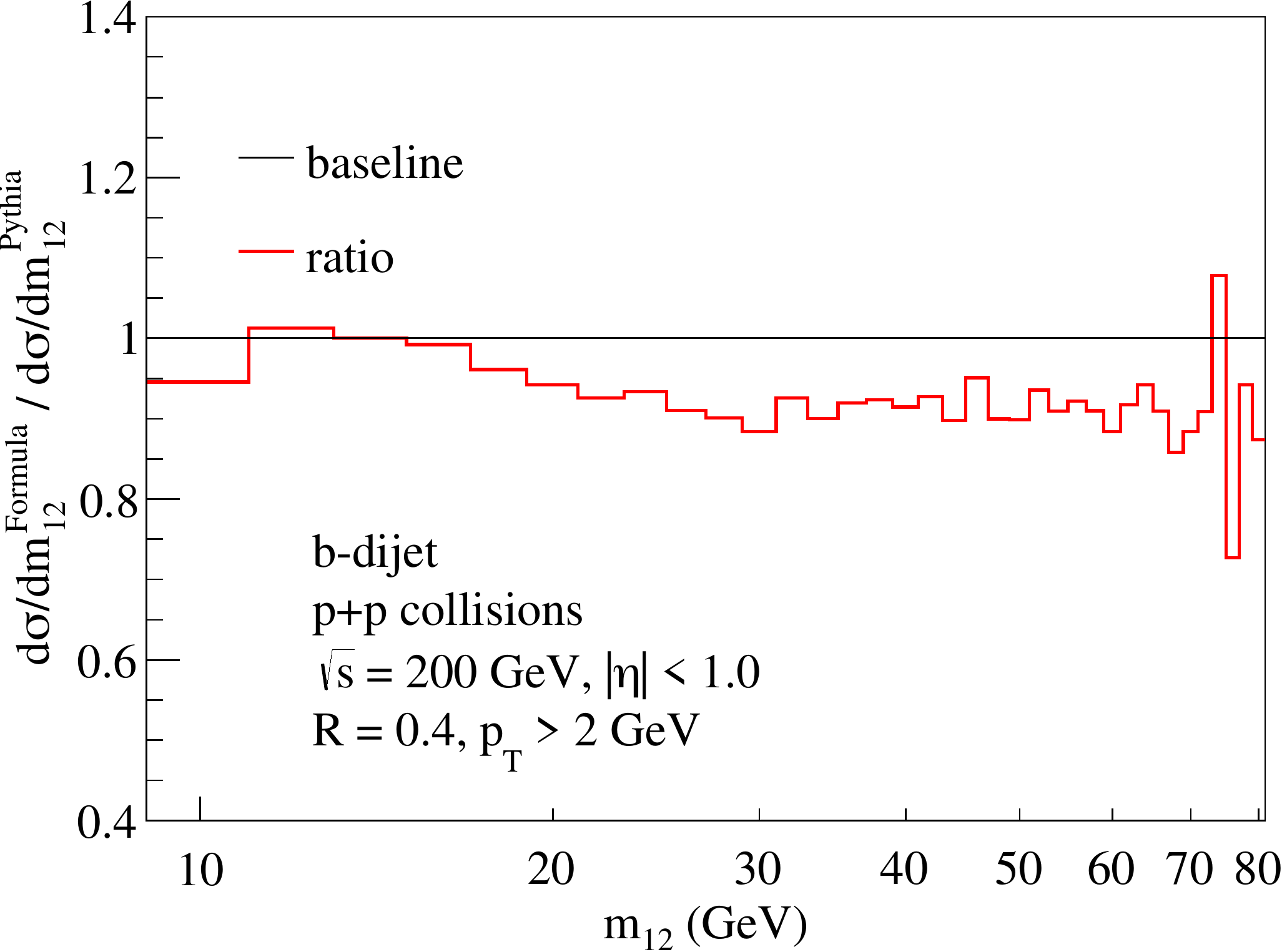}
\hskip 0.2in
\includegraphics[width=3.0in]{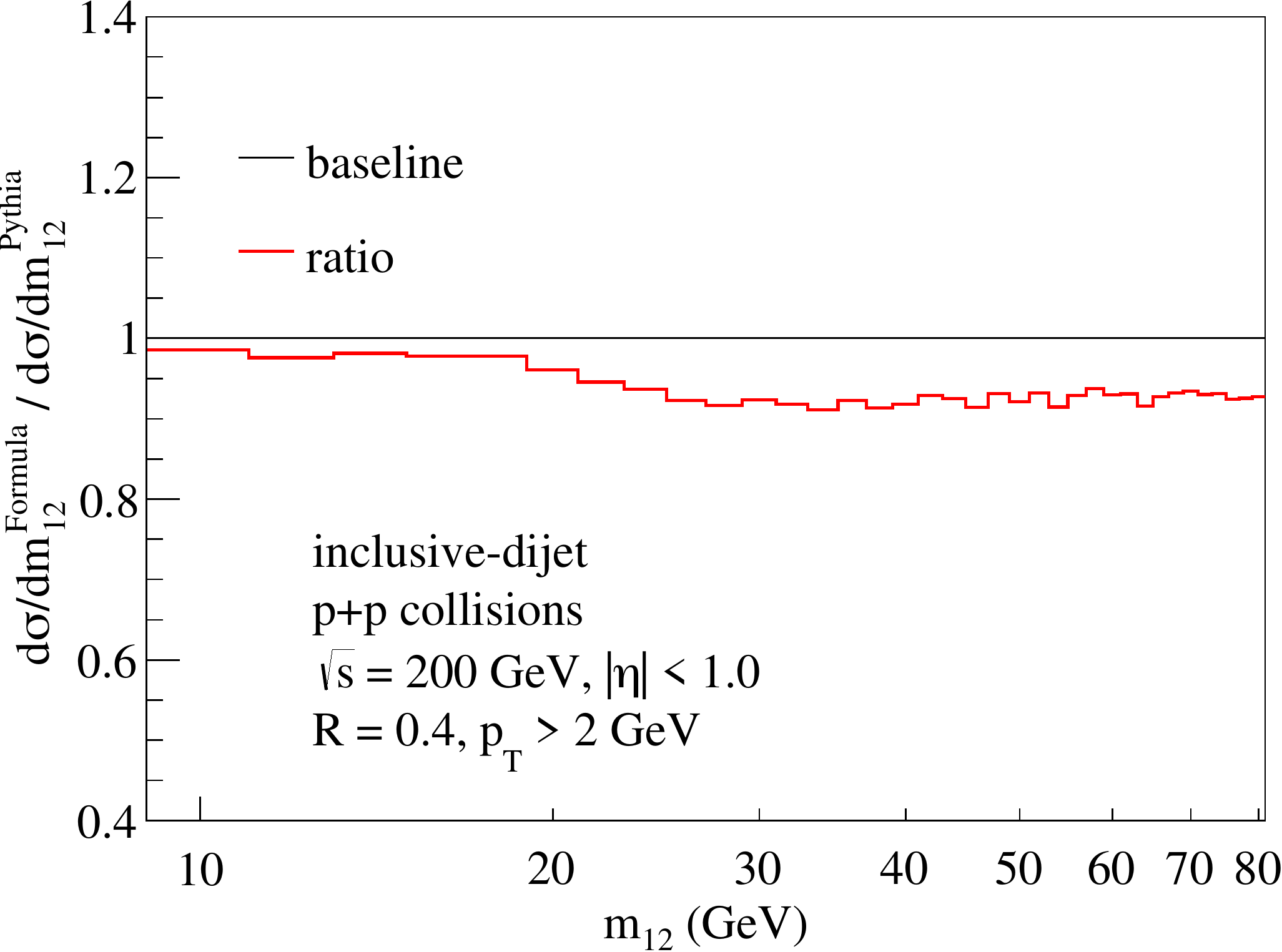}
\caption{Mass distributions (top) and their ratios (bottom) for $b$-tagged (left) and inclusive (right) dijet production in p+p collisions at $\sqrt{s}=200$ GeV. Kinematic cuts implemented in our simulations are the same as those from the sPHENIX collaboration~\cite{sPHENIX}. The upper panels display black histograms representing $d\sigma/dm_{12}$ simulated {\it directly} from Pythia 8, while the red histograms are $d\sigma/dm_{12}$ computed using Eq.~\eqref{eq:mass} and Pythia 8-simulated~$d\sigma/dp_{1T}dp_{2T}$. In the lower panels the Pythia calculation is given by the black lines and the red histograms represent the ratio  of the approximate formula to the baseline.}
\label{fig:mass_check_sPHENIX}
\eef

Let us now confirm that such a procedure yields correct dijet mass distributions. To do this, we compare the dijet invariant mass distribution {\it indirectly} computed using Eq.~\eqref{eq:mass} and Pythia 8-simulated $d\sigma/dp_{1T}dp_{2T}$, with $d\sigma/dm_{12}$ simulated {\it directly} from Pythia 8. We perform such a comparison in Figs.~\ref{fig:mass_check_LHC} and \ref{fig:mass_check_sPHENIX} for $b$-tagged (left panels), as well as for inclusive (right panels), dijet production at LHC energy $\sqrt{s} = 5.02$ TeV and sPHENIX energy $\sqrt{s} = 200$ GeV, respectively. In the top panels, the black histograms represent $d\sigma/dm_{12}$ simulated {\it directly} from Pythia 8, while the red histograms are $d\sigma/dm_{12}$ computed using Eq.~\eqref{eq:mass} and Pythia 8-simulated $d\sigma/dp_{1T}dp_{2T}$. In the bottom panels, the black histograms mark the baseline at unity for the mass distribution ratios while the red histograms represent the ratio between the mass distributions utilizing Eq.~\eqref{eq:mass} and those directly from Pythia 8. We observe a quite reasonable matching of mass spectra obtained via direct implementation of dijet mass in Pythia 8 and our approximate formula in Eq.~\eqref{eq:mass}, as indicated by the fact that the approximate distributions only deviates by  $ \leq 10\%$ from the exact simulation. This induces a minor change in the overall normalization of each distribution whose effect cancels out in the computation of the nuclear modification factor $R_{AA}$. This validates the use of our formula in applications to heavy-ion collisions and subsequent dijet mass modifications.

On the other hand, one of the more conventional observables, the dijet momentum imbalance shift, is based on the cross section as a function of the imbalance variable 
\bea
z_J = p_{2T}/p_{1T},
\eea 
which can be derived from the double differential cross section $d\sigma/dp_{1T}dp_{2T}$. The formula is given as follows
\bea
\frac{d\sigma}{dz_J} = \int dp_{1T} dp_{2T} \frac{d\sigma}{dp_{1T} dp_{2T}} \delta\left(z_J - \frac{p_{2T}}{p_{1T}}\right),
\label{eq:zJ}
\eea
where again, the limits of integration for $p_{1T}$ and $p_{2T}$ are matched to the desired experimental cuts. 

Comparing Eqs.~\eqref{eq:mass} with \eqref{eq:zJ}, one can gain some insights why medium modification of dijet invariant mass distribution leads to enhanced medium effects than that of the dijet momentum imbalance. This is because dijet invariant mass $m_{12}\propto p_{1T}p_{2T}$, i.e. product of two jet momenta, and thus leads to a combination of the jet quenching effects on the individual jets. On the other hand, the momentum imbalance $z_J = p_{2T}/p_{1T}$, i.e. quotient of two jet momenta, and thus diminishes the jet quenching effects on the individual jets. We elaborate more on this point in the presentation of our numerical results below. 

\subsection{Modification of dijet production}
In the presence of the hot and dense QCD medium, the vacuum parton shower gets modified due to the radiative~\cite{Zakharov:1997uu,Baier:1996kr,Gyulassy:2000fs,Wiedemann:2000za,Wang:2001ifa,Arnold:2002ja,Zhang:2003wk,Dokshitzer:2001zm} and collisional~\cite{Braaten:1991we,Wicks:2005gt,Adil:2006ei,Thoma:2008my,Berrehrah:2013mua,Neufeld:2014yaa} energy losses of the propagating partons that initiate and form the jets. The implementation of energy loss effects in heavy ion collisions is explained in detail in, e.g., Refs.~\cite{He:2011pd,Kang:2017xnc}. For a given impact parameter $|{\bf b}_\perp |$ in the transverse plane of the nucleus-nucleus collisions, we evaluate the inclusive dijet double differential cross sections in~$(p_{1T}, p_{2T})$ as follows
\bea
\frac{d\sigma^{AA}(|{\bf b}_\perp |) }{dp_{1T}dp_{2T}} 
=&  \int d^2 {\bf s_\perp}   T_A\left( {\bf s}_\perp - \frac{{\bf b}_\perp}{2}  \right)
T_A\left( {\bf s}_\perp + \frac{{\bf b}_\perp}{2}  \right)  
\nonumber \\
&\times 
\sum_{q,g} 
\int_0^1 d\epsilon \frac{P_{q,g}^1(\epsilon; {\bf s}_\perp, |{\bf b}_\perp | )}{1-f_{q,g}^{\rm 1\; loss}(R;s_\perp, |{\bf b}_\perp | )\, \epsilon}
\int_0^1d\epsilon' \frac{P_{q,g}^2(\epsilon'; {\bf s}_\perp, |{\bf b}_\perp | )}{1-f_{q,g}^{\rm 2\; loss}(R;s_\perp, |{\bf b}_\perp | )\, \epsilon'}
\nonumber \\
&\times
\frac{d\sigma^{NN}_{q,g}  \left( p_{1T}/ [1-f_{q,g}^{\rm 1\,  loss}(R; s_\perp, |{\bf b}_\perp | )\,  \epsilon], 
p_{2T}/ [1-f_{q,g}^{\rm2\,  loss}(R; s_\perp, |{\bf b}_\perp | )\,  \epsilon'] \right)}{dp_{1T}dp_{2T}}\, , 
\label{eq:master}
\eea
where $|{\bf b}_\perp |$ is the mean impact parameter for a given collision centrality.
For the $b$-tagged dijet case, we further include the contributions from $b$-quarks. 
In Eq.~\eqref{eq:master}, $T_A\left({\bf s}_\perp  \right) = \int_{-\infty}^{\infty}   \rho_A({\bf s}_\perp,z ) dz$ is the so-called thickness function in the usual optical Glauber model, where we choose the inelastic nucleon-nucleon scattering cross section $\sigma_{\rm in} = 70$ mb ($42$ mb) to obtain average number of binary collisions at $\sqrt{s_{NN}} =5.02 $~TeV (200~GeV)~\cite{Miller:2007ri}, respectively. $ P_{q,g}(\epsilon) $ is the probability density for the parent parton to redistribute a fraction $\epsilon$ of its energy through medium-induced soft gluon bremsstrahlung. For reconstructed jets, what matters is the out-of-cone energy loss fraction $f_{q,g}^{\rm loss}$~\cite{Kang:2017xnc} 
\bea
f_{q,g}^{\rm loss}(R;{\rm rad+coll}) = 1- \left( \int_0^{R }  dr  \int_{\omega_{\rm min}}^E  d\omega \,     \frac{dN^g_{q,g}(\omega,r)}{d\omega dr}    \right)  \Bigg/
 \left(  \int_0^{R_{\rm max} }   dr  \int_0^E  d\omega \,     \frac{dN^g_{q,g}(\omega,r)}{d\omega dr}   \right) \;,
\eea
which includes both radiative and collisional energy loss effects, with $\omega_{\rm min}$ being a parameter that controls the energy dissipated by the medium-induced parton shower into the QGP due to collisional processes~\cite{Neufeld:2014yaa}. On the other hand, 
$\frac{dN^g_{q,g}(\omega,r)}{d\omega dr}$ is the medium-induced gluon distribution~\cite{Vitev:2007ve}, which is the soft emission limit of the complete in-medium splitting functions~\cite{Ovanesyan:2011kn}.

Splitting functions themselves are calculated using the formula derived in the $\textrm{SCET}_\textrm{M,G}$ framework \cite{Kang:2016ofv}. They have been independently obtained in the lightcone wavefunction approach~\cite{Sievert:2018imd} for both massless and massive partons, and are evaluated in the QGP medium simulated by the iEBE-VISHNU code package \cite{Shen:2014vra}. The same model of the medium has been recently used to calculate quarkonium suppression at the LHC~\cite{Aronson:2017ymv} and soft-drop groomed momentum sharing distributions~\cite{Li:2017wwc}. Numerical evaluation of the splitting functions requires multi-dimensional integration over the jet production point, the propagation of the jet in matter, and the transverse momentum dependence of the jet-medium cross section. Since the integral dimension is larger than 4, we use a numerical integration based on the Monte Carlo method. In particular, the VEGAS algorithm \cite{PETERLEPAGE1978192} implemented in the CUBA multidimensional numerical integration library \cite{Hahn:2004fe} is used because the adaptive importance sampling algorithm is efficient for integrands with localized peaks. The splitting function calculation code is written in C++. The integrals are evaluated on a Xeon cluster with task parallelization for different kinematic variables such as energy, momentum, quark mass, or the splitting channel, utilizing multiple CPU cores. Integration ranges are determined following the study presented in Ref.~\cite{Ovanesyan:2011kn}.

Once we obtain the medium-modified differential cross section $d\sigma^{AA}/dp_{1T}dp_{2T}$, we then use Eqs.~\eqref{eq:mass} and \eqref{eq:zJ} to compute the dijet invariant mass distribution $d\sigma^{AA}/dm_{12}$ and imbalance distribution $d\sigma^{AA}/dz_J$ in heavy ion collisions. Such a procedure is perfectly fine for $d\sigma^{AA}/dz_J$, but is an approximation for $d\sigma^{AA}/dm_{12}$, where we assume that the medium modification for the single jet mass distributions $\langle m_1^2\rangle$ and $\langle m_2^2\rangle$ are much smaller than those for the transverse momenta $p_{1T}$ and $p_{2T}$. Thus, starting from Eq.~\eqref{eq:mass}, we obtain
\bea
\frac{d\sigma^{AA}}{dm_{12}} = \int dp_{1T} dp_{2T} \frac{d\sigma^{AA}}{dp_{1T}dp_{2T}} \delta\left(m_{12} - \sqrt{\langle m_1^2\rangle_{pp} + \langle m_2^2\rangle_{pp} + 2p_{1T}p_{2T}\langle\mathrm{cosh(\Delta \eta)} - \mathrm{cos}(\Delta \phi)\rangle_{pp}} \right),
\label{eq:massAA}
\eea
where we have used the same values for $\langle m_1^2\rangle$, $\langle m_2^2\rangle$, and $\langle\mathrm{cosh(\Delta \eta)} - \mathrm{cos}(\Delta \phi)\rangle$ as those in p+p collisions, as denoted by the subscript $pp$. Such an approximation is well-justified. For example, mass distributions for single inclusive jets are indeed not significantly modified, as observed by ALICE collaboration at the LHC~\cite{Acharya:2017goa}.

\section{Phenomenological results at RHIC and the LHC}
In this section we first present our phenomenological results for both inclusive and $b$-tagged dijet production in A+A collisions at the LHC, as well as the future sPHENIX experiment at RHIC. To investigate dijet production in heavy ion collisions and quantify its deviation from the baseline results in elementary p+p reactions, we start with the two-dimensional nuclear modification factor
\bea
R_{AA}(p_{1T}, p_{2T}, |{\bf b}_\perp|) = \frac{1}{\langle N_{\rm bin}\rangle}\frac{d\sigma^{AA}(|{\bf b}_\perp|)/dp_{1T}dp_{2T}}{d\sigma^{pp}/dp_{1T}dp_{2T}},
\eea
where $|{\bf b}_\perp|$ is the corresponding impact parameter and $\langle N_{\rm bin}\rangle$ is the average number of nucleon-nucleon scatterings for a given centrality class. In this paper, we focus on the most central collisions. In Fig.~\ref{fig:RAA_3D_LHC}, we make 3D plots for nuclear modification factor $R_{AA}$ as a function of the jet transverse momenta $p_{1T}$ and $p_{2T}$ simultaneously. The calculations are done for the production of dijets with radii $R=0.4$ in central ($0-10\%$) Pb+Pb collisions at the LHC energy $\sqrt{s_{NN}} = 5.02$ TeV. We integrate the rapidities of both jets over the interval $|y|<2$. For the medium effects, we choose the coupling between the jet and the medium to be $g=1.8$. This is consistent with the value used in our previous studies for single inclusive jets~\cite{Kang:2017frl}, vector-boson-tagged jets~\cite{Kang:2017xnc}, jet substructure~\cite{Chien:2015hda,Chien:2016led}, and single inclusive hadrons~\cite{Kang:2014xsa,Chien:2015vja,Kang:2016ofv} in A+A collisions. The left figure is for $b$-tagged dijet production, while the right is for inclusive dijets. We note that while we plot the full symmetric range in $p_{1T}$ and $p_{2T}$, we do have in mind that the first jet (1) will be the trigger or leading jet and the second jet (2) will be the recoil or subleading jet. Thus, we incorporate on average path length and color charge bias effects in our calculation.

\begin{figure}[hbt]
\includegraphics[width=3.0in]{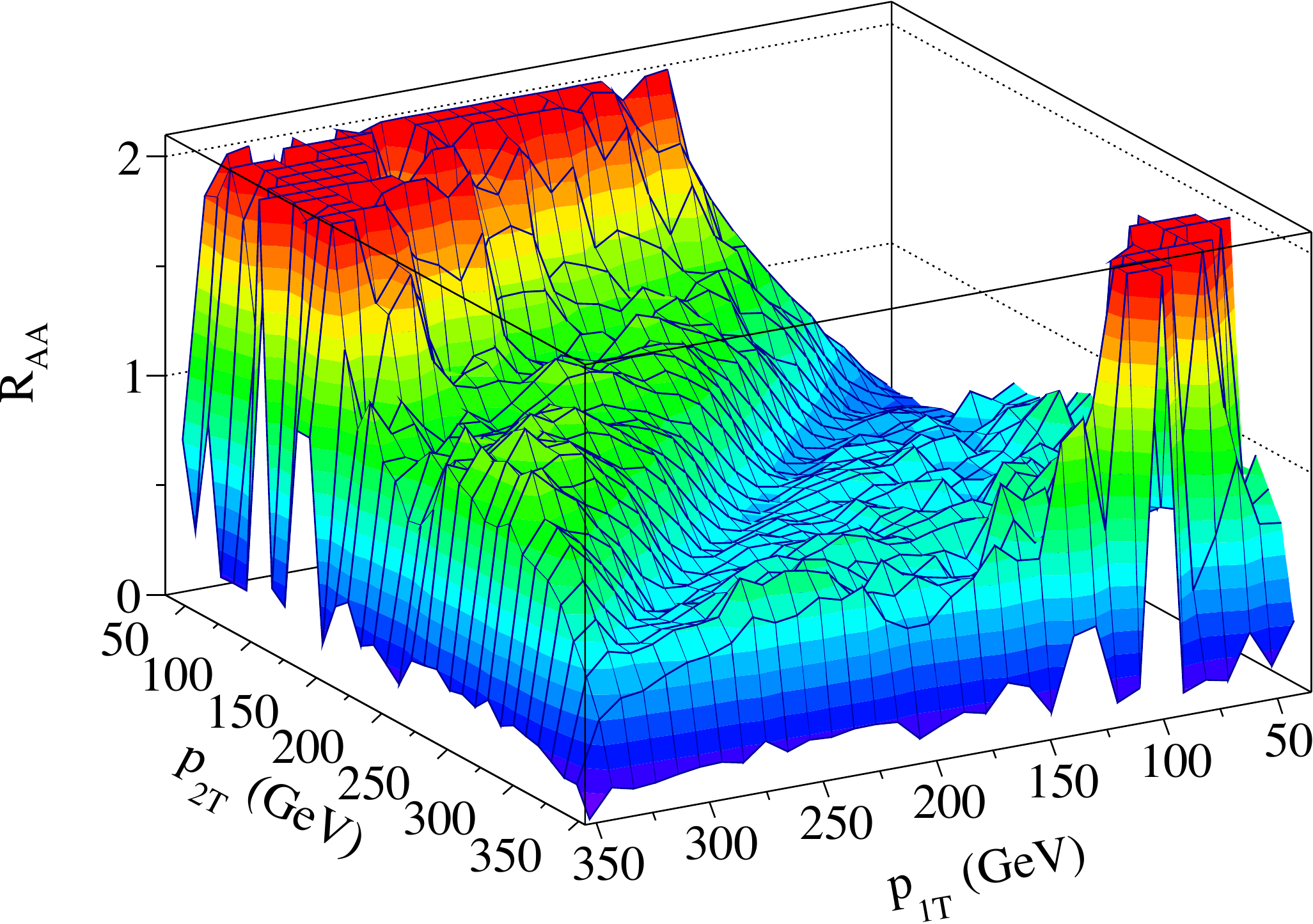}
\hskip 0.2in
\includegraphics[width=3.0in]{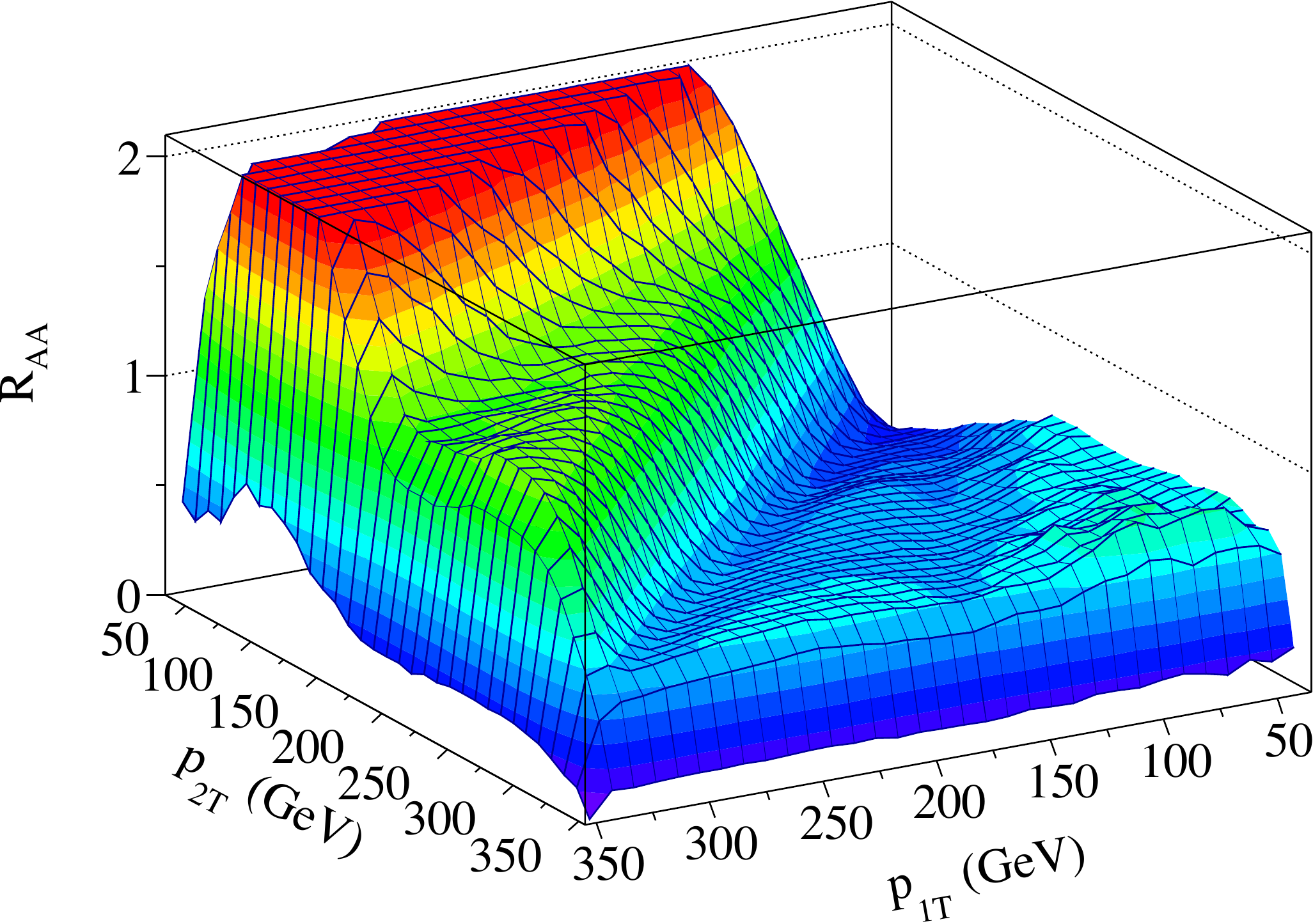}
\caption{Nuclear modification factor for $b$-tagged (left) and inclusive (right) dijet production in p+p collisions at $\sqrt{s}=5.02$ TeV. Kinematic cuts are implemented in our simulations as in CMS measurements~\cite{Sirunyan:2018jju}.}
\label{fig:RAA_3D_LHC}
\eef

\begin{figure}[hbt]
\includegraphics[width=3.0in]{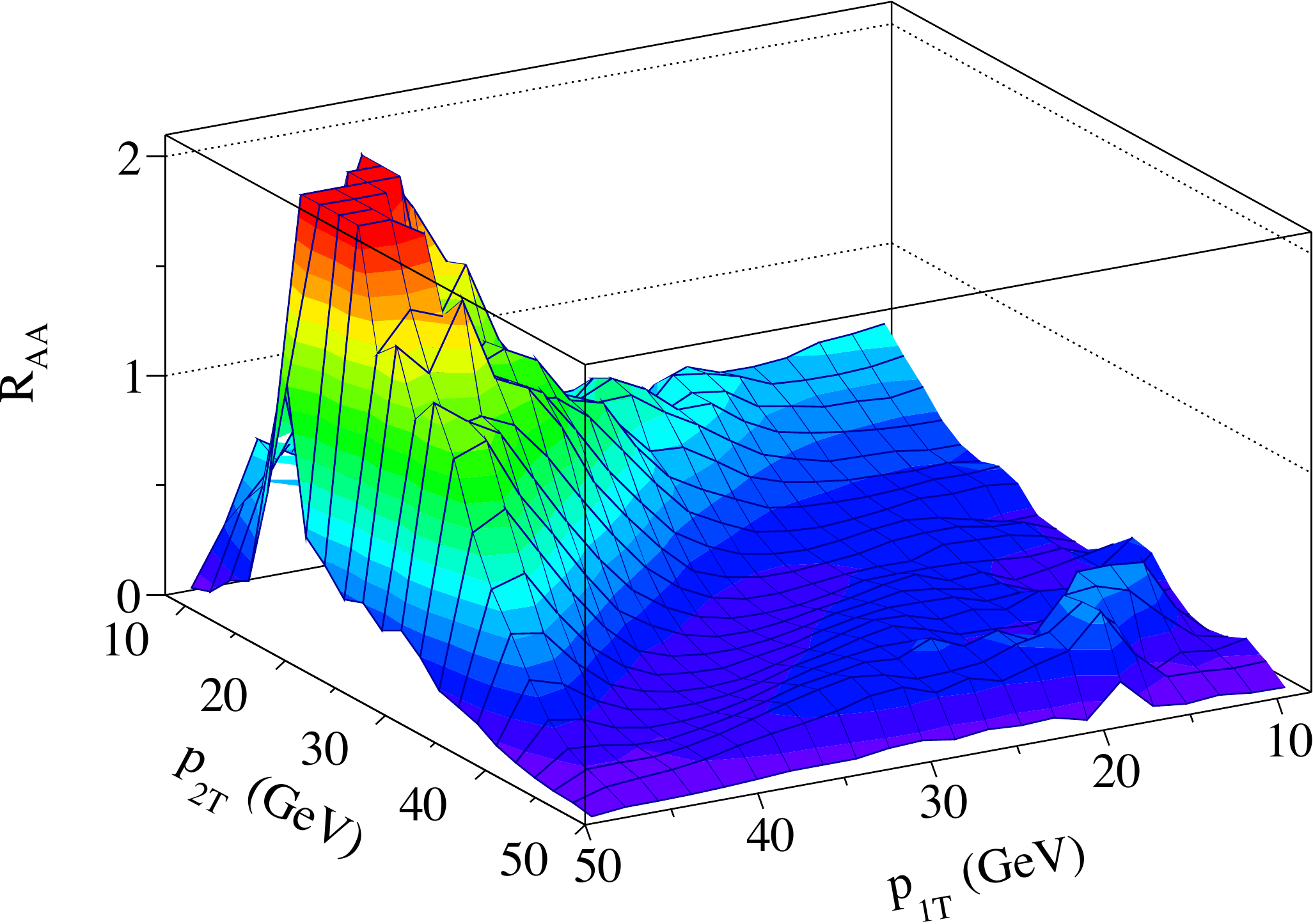}
\hskip 0.2in
\includegraphics[width=3.0in]{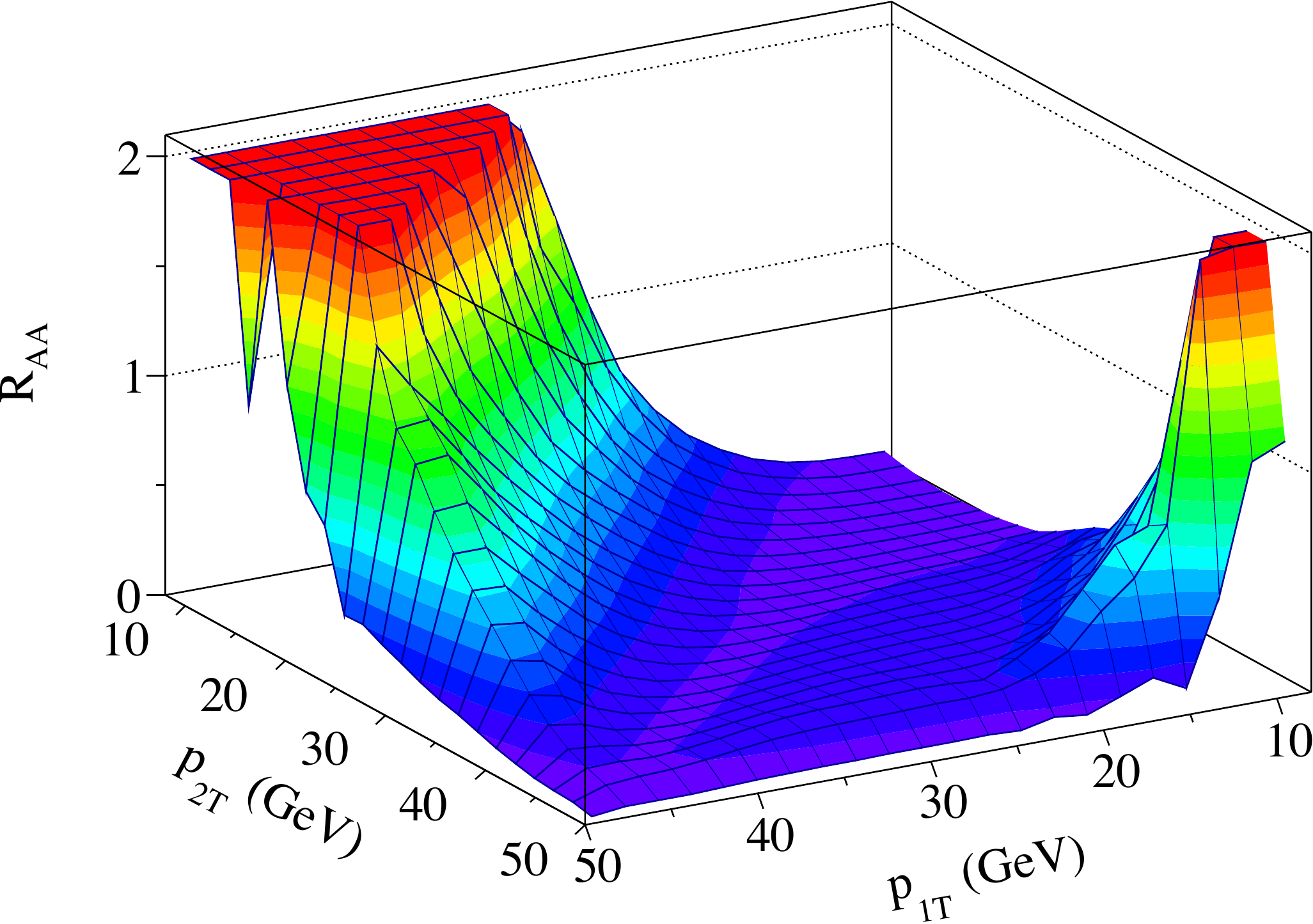}
\caption{Nuclear modification factor for $b$-tagged (left) and inclusive (right) dijet production in p+p collisions at $\sqrt{s}=200$ GeV. Kinematic cuts implemented in our simulations are the same as those from the sPHENIX collaboration~\cite{sPHENIX}.}
\label{fig:RAA_3D_sPHENIX}
\eef

As one can clearly see, the largest suppression occurs along the diagonal $p_{1T} = p_{2T}$, consistent with our expectation. In the region away from the diagonal, there is a striking enhancement. As the future sPHENIX~\cite{Adare:2015kwa} experiment will have good sensitivity in measuring both inclusive and $b$-tagged dijet production, it is an opportune time to make predictions for sPHENIX kinematics. In Fig.~\ref{fig:RAA_3D_sPHENIX} we make similar 3D plots of $R_{AA}$ for $b$-tagged (left) and inclusive (right) dijet production at sPHENIX energy $\sqrt{s_{NN}} = 200$ GeV. Kinematic cuts implemented in our simulations are the same as those from the sPHENIX collaboration~\cite{sPHENIX}. Obviously the kinematic coverage for the jet transverse momenta is much smaller than that of the jets at the LHC, due to a much smaller center-of-mass energy. However, the suppression is even stronger along the diagonal $p_{1T} = p_{2T}$. This is simply because the cross sections at RHIC energies fall much faster as functions of jet transverse momenta due to limited phase space, and thus jet quenching effects get amplified~\cite{Vitev:2004gn,Adil:2004cn,Wang:2004tt,Abelev:2007ra,Adare:2008ad}.

If such two-dimensional nuclear modification ratios could be measured in detail, they would provide the most information and insight into jet quenching and heavy flavor dynamics in the medium. However, the statistics necessary to perform such measurements make this, at present, quite difficult. In practice, one usually integrates out one of the differential variables and, thus, achieves a one-dimensional nuclear modification ratio. In this respect, the conventional dijet momentum imbalance $z_J$ and asymmetry $A_J$ distributions have been extensively studied in the literature. The medium modification on these traditional distributions emphasize the difference in the quenching of the dijet production, which has been observed to be relatively small. We will present such studies toward the end of this section.

\bef
\includegraphics[width=3.0in]{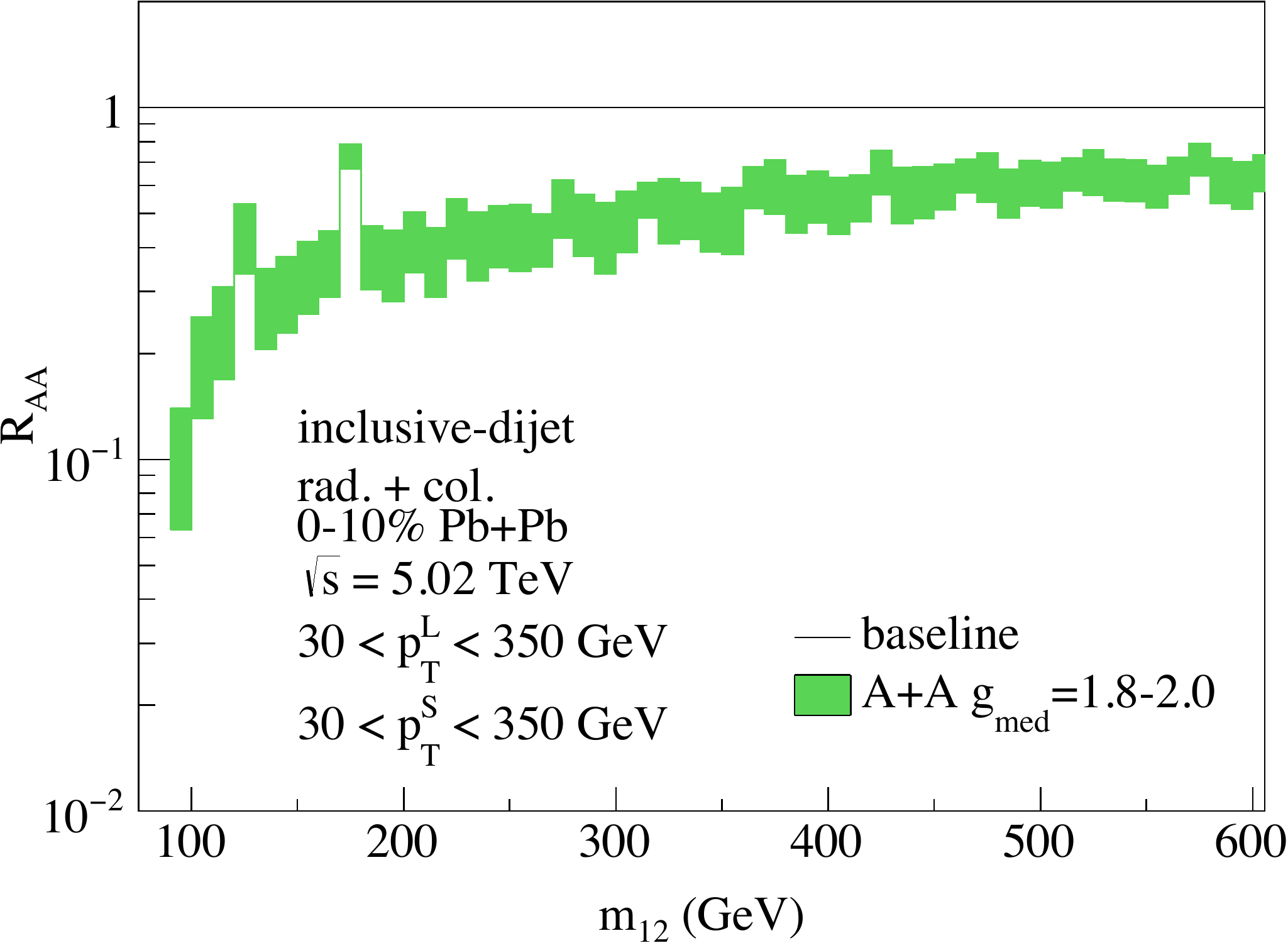}
\hskip 0.2in
\includegraphics[width=3.0in]{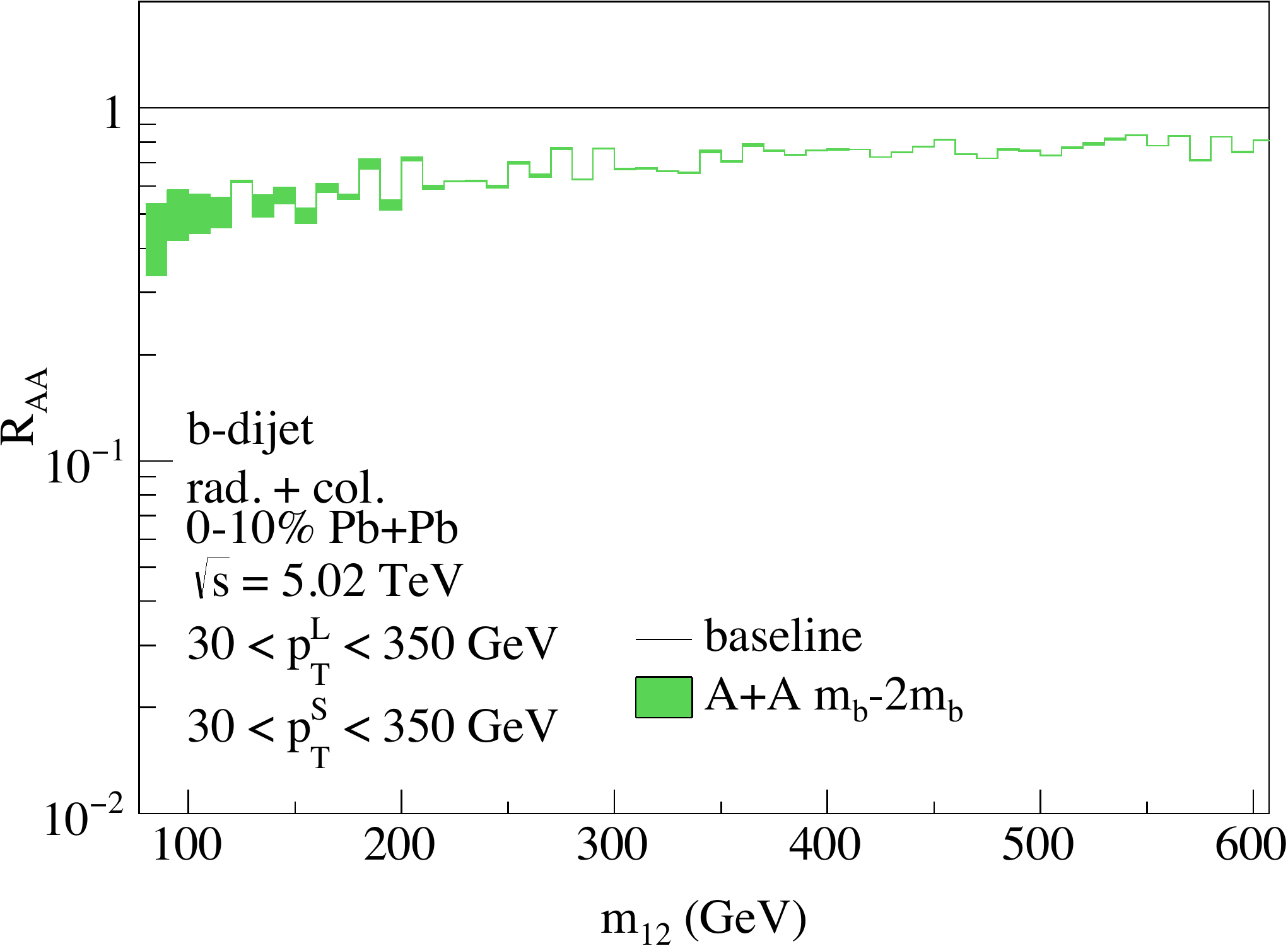}
\caption{Nuclear modification factor $R_{AA}$ is plotted as a function of dijet invariant mass $m_{12}$ for inclusive (left) and $b$-tagged (right) dijet production in Pb+Pb collisions at $\sqrt{s_{NN}} = 5.02$ TeV at the LHC. Left: the band corresponds to a range of coupling strength between the jet and the medium: $g_{\rm med}=1.8-2.0$, respectively. Right: we fix $g_{\rm med}=1.8$, and the band corresponds to a range of masses of the propagating system between $m_b$ and $2m_b$.}
\label{fig:CMS-mass}
\eef

Here instead, we present the nuclear modification for another observable, the dijet invariant mass distribution, defined as follows
\bea
R_{AA}(m_{12}, |{\bf b}_\perp|) = \frac{1}{\langle N_{\rm bin}\rangle}\frac{d\sigma^{AA}(|{\bf b}_\perp|)/dm_{12}}{d\sigma^{pp}/dm_{12}}.
\eea
Again, the impact parameter $|{\bf b}_\perp|$ indicates the centrality class for the A+A collisions. The numerator and denominator are the dijet mass distribution in A+A and p+p collisions, respectively. They are computed through the double differential cross sections $d\sigma/d_{1T}dp_{2T}$ as in Eqs.~\eqref{eq:massAA} and \eqref{eq:mass}, respectively. In Eqs.~\eqref{eq:mass} and \eqref{eq:massAA}, one can immediately see the advantage of such an observable. First, being only differential in the dijet invariant mass $m_{12}$, it is a one-dimensional observable, hence one should have enough statistics to perform these measurements experimentally. Second, since the dijet invariant mass is proportional to the product of the dijet transverse momenta, as can be clearly seen in Eq.~\eqref{eq:massform}, the dijet mass distribution incorporates the medium modification of the $d\sigma/d_{1T}dp_{2T}$ in an amplified way, as emphasized in Sec.~\ref{sec:main formula}. In other words, compared to the traditional momentum asymmetry observables, the dijet mass distribution {\it combines} rather than subtracts the medium modifications of the two jets. Naturally, one would expect the medium modification of dijet mass distributions to be greatly enhanced and thus to be more sensitive to the properties of the medium.

\bef
\includegraphics[width=3.0in]{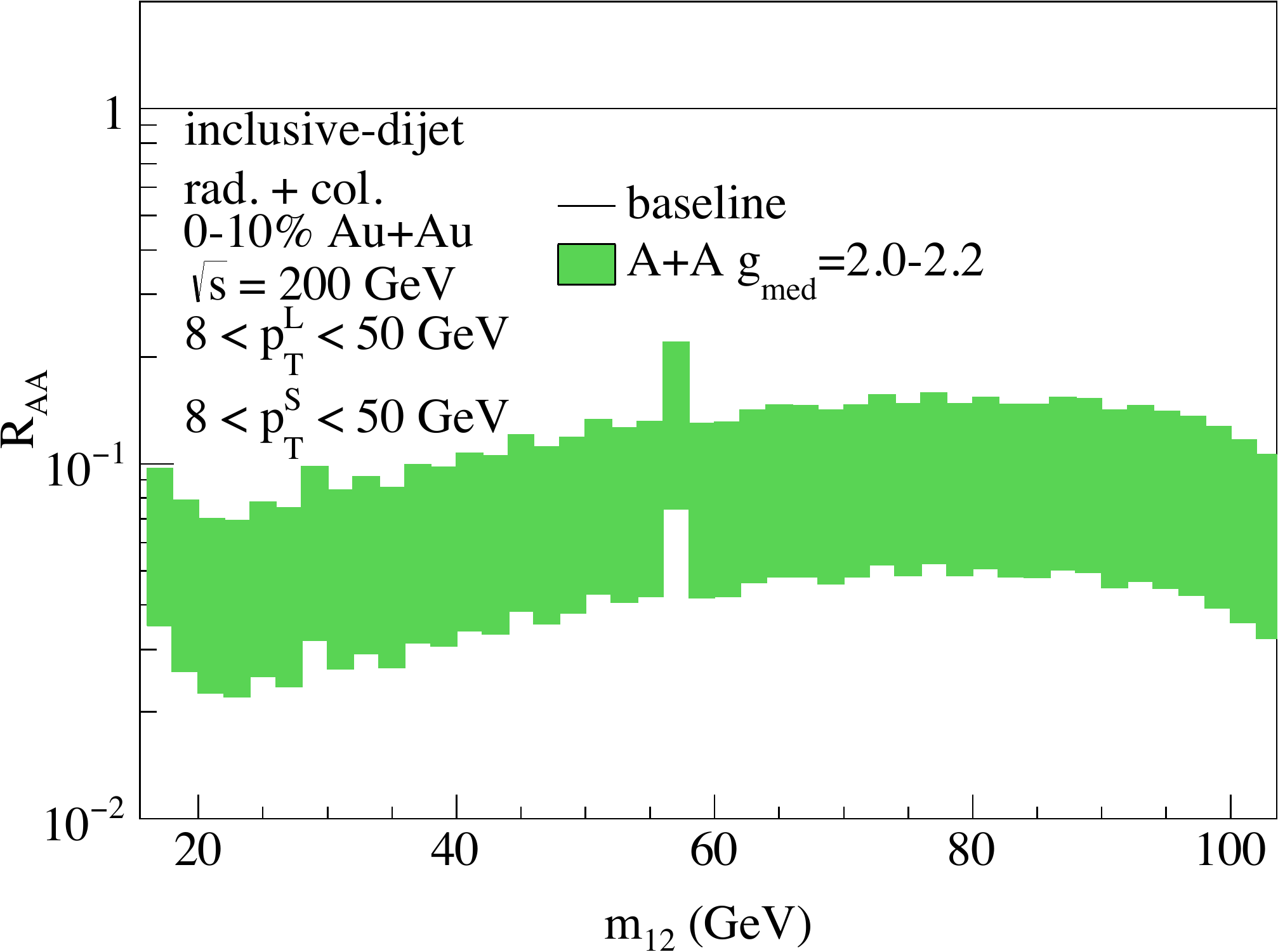}
\hskip 0.2in
\includegraphics[width=3.0in]{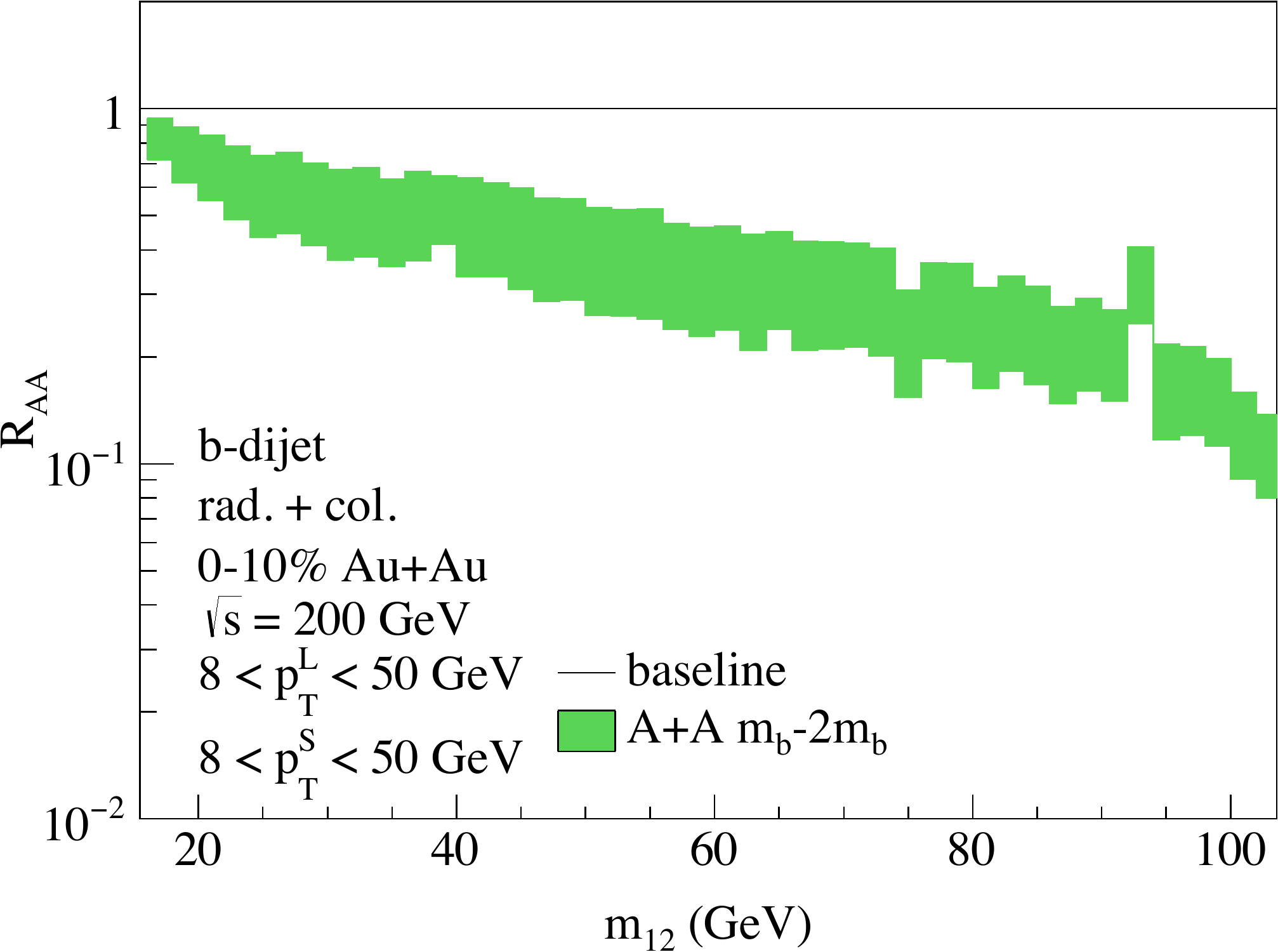}
\caption{Nuclear modification factor $R_{AA}$ plotted as a function of dijet invariant mass $m_{12}$ for inclusive (left) and $b$-tagged (right) dijet production in Au+Au collisions at $\sqrt{s_{NN}} = 200$ GeV for sPHENIX at RHIC. Left: the band corresponds to a range of coupling strength between the jet and the medium: $g_{\rm med}=2.0-2.2$, respectively. Right: we fix $g_{\rm med}=2.0$, and the band corresponds to a range of masses of the propagating system between $m_b$ and $2m_b$.}
\label{fig:phenix-mass}
\eef

In Fig.~\ref{fig:CMS-mass}, we plot the nuclear modification factor $R_{AA}$ as a function of dijet invariant mass $m_{12}$ for inclusive (left) and $b$-tagged (right) dijet production in Pb+Pb collisions at $\sqrt{s_{NN}} = 5.02$ TeV at the LHC. For inclusive dijet production, the band corresponds to a range of coupling strengths between the jet and the medium: $g_{\rm med}=1.8-2.0$. On the other hand, for $b$-tagged dijet production, we fix $g_{\rm med}=1.8$, and the band corresponds to a range of masses of the propagating system between $m_b$ and $2m_b$, implemented as detailed in~\cite{Huang:2013vaa}. We make transverse momentum cuts requiring both leading and subleading jets to have $p_T^{\rm L, S} > 30$ GeV. This is why we have a lower limit on the dijet invariant mass $m_{12}\gtrsim 100$ GeV in these plots. As one can clearly see from the figures, being an amplifying effect, $R_{AA}$ can be as small as 0.1, i.e., suppressed by a factor of 10 in the lower end of the invariant mass $m_{12}\sim 100$ GeV. This is a dramatic suppression, much stronger than the suppression for single inclusive jet production, around a factor of 2~\cite{Kang:2017frl}. As one increases the invariant mass $m_{12}$, the suppression gets smaller, but it is still around a factor of 2 or more even at $m_{12}\sim 500$ GeV. The suppression for $b$-tagged dijet production is smaller than that of inclusive dijets at smaller dijet mass $m_{12}\sim 100$ GeV, and becomes similar to inclusive dijet production as $m_{12}$ increases. This is to be expected, as heavy quark mass effects on jet quenching are more important at lower transverse momenta, or naturally smaller dijet invariant mass.

\begin{figure}[hbt]
\includegraphics[width=3.0in]{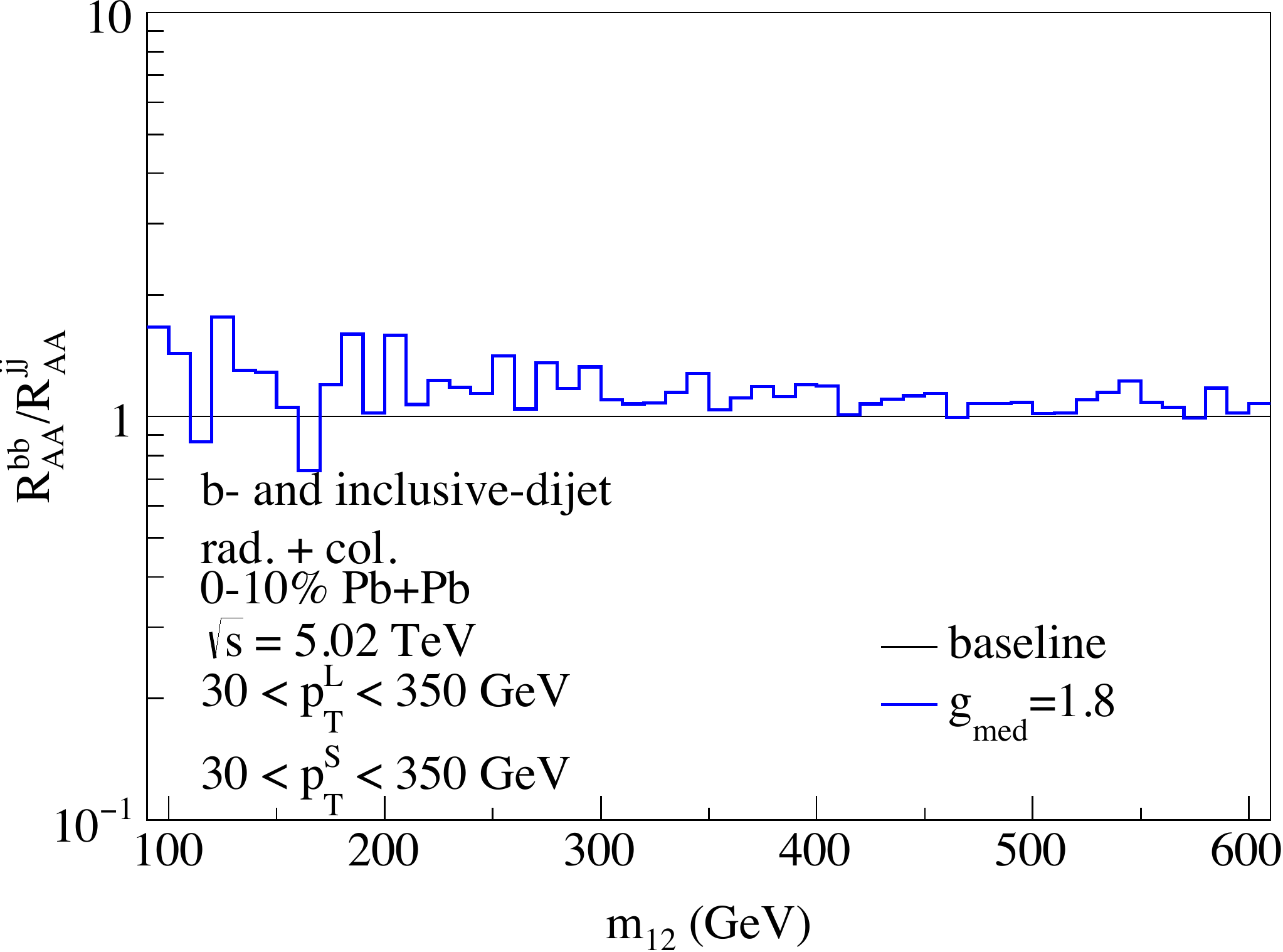}
\hskip 0.2in
\includegraphics[width=3.0in]{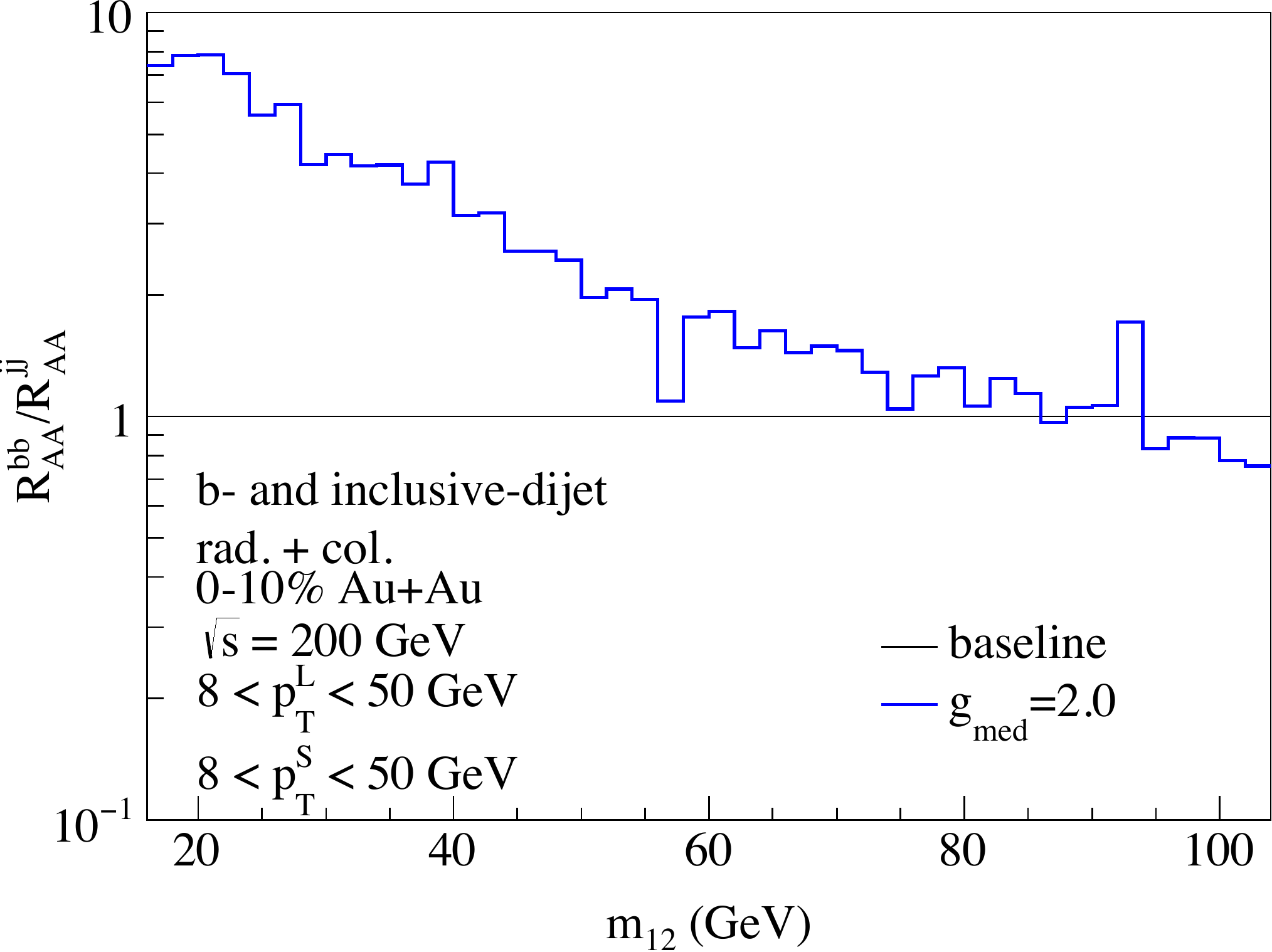}
\caption{Ratios of nuclear modification factors for $b$-tagged ($R_{AA}^{bb}$) v.s inclusive ($R_{AA}^{jj}$) dijet production for CMS (left) and sPHENIX (right) are plotted as a function of dijet invariant mass $m_{12}$. For LHC (sPHENIX) energies, we choose $g_{\rm med} = 1.8~(2.0)$. For $b$-tagged dijets, the mass of the propagating system is held fixed at $m_b$.}
\label{fig:R_AA_mass_ratio}
\eef

In Fig.~\ref{fig:phenix-mass}, we present the same plots but for Au+Au collisions at $\sqrt{s_{NN}} = 200$ GeV, relevant to the sPHENIX experiment at RHIC. For inclusive dijet production, the band corresponds to a range of coupling strengths between the jet and the medium: $g_{\rm med}=2.0-2.2$. On the other hand, for $b$-tagged dijet production, we fix $g_{\rm med}=2.0$, and the band again corresponds to a range of masses of the propagating system between $m_b$ and $2m_b$. We choose a slightly larger coupling strength at RHIC compared to that for the above LHC kinematics, which is also consistent with our previous studies and that of the JET collaboration~\cite{Burke:2013yra}. Since the center-of-mass energy is much lower, we select jets with much lower $p_T \gtrsim 8 $~GeV, and correspondingly lower dijet invariant mass $m_{12} \gtrsim 20$ GeV for RHIC kinematics. Having smaller jet transverse momenta and cross sections that fall off strongly as functions of jet transverse momenta, the suppression for inclusive dijet cross sections is even larger compared with those of LHC energies. We observe a factor of $\sim 10$ or more suppression even up to a relatively high invariant mass $m_{12}\sim 100$ GeV.

On the other hand, the suppression pattern for $b$-tagged dijet production as a function of $m_{12}$ at sPHENIX energy $\sqrt{s_{NN}} = 200$ GeV, as shown in right panel of Fig.~\ref{fig:phenix-mass}, appears quite different from inclusive dijet production in left panel, and looks nothing like the $b$-tagged dijet production at the LHC energy in Fig.~\ref{fig:CMS-mass}. It is, thus, important to understand why we observe such a behavior. If one recalls the behavior of the suppression pattern for single inclusive heavy meson/heavy quark production as a function of its transverse momentum, see, e.g. Ref.~\cite{Kang:2016ofv,Cao:2018ews}, one can understand the above behavior of $R_{AA}$ as a function of $m_{12}$. Due to the heavy quark mass effect in the jet quenching formalism, $R_{AA}$ for heavy quark mesons first decreases and then increases when plotted as a function of $p_T$. In other words, there is a dip in $R_{AA}$ as a function of $p_T$. Now one can translate such a behavior into the behavior of $R_{AA}$ as a function of $m_{12}$. For the mass region in Fig.~\ref{fig:phenix-mass}, $b$-tagged dijets mostly fall into the relatively low values of jet transverse momenta, i.e., before the dip of $R_{AA}$ (as a function of $p_T$). This explains why $R_{AA}$ decreases as a function of $m_{12}$. If one has a larger phase space to explore much higher values of transverse momenta, as is the case at the LHC energy in Fig.~\ref{fig:CMS-mass}, once passing the dip of $R_{AA}$, one should naturally expect $R_{AA}$ to increase as a function of $m_{12}$. This is precisely what is observed in our calculations, see Fig.~\ref{fig:CMS-mass} (right). This comparison informs us that sPHENIX is sitting in a very interesting kinematic regime for testing heavy quark mass effects within the jet quenching formalism.

\bef
\includegraphics[width=3.0in]{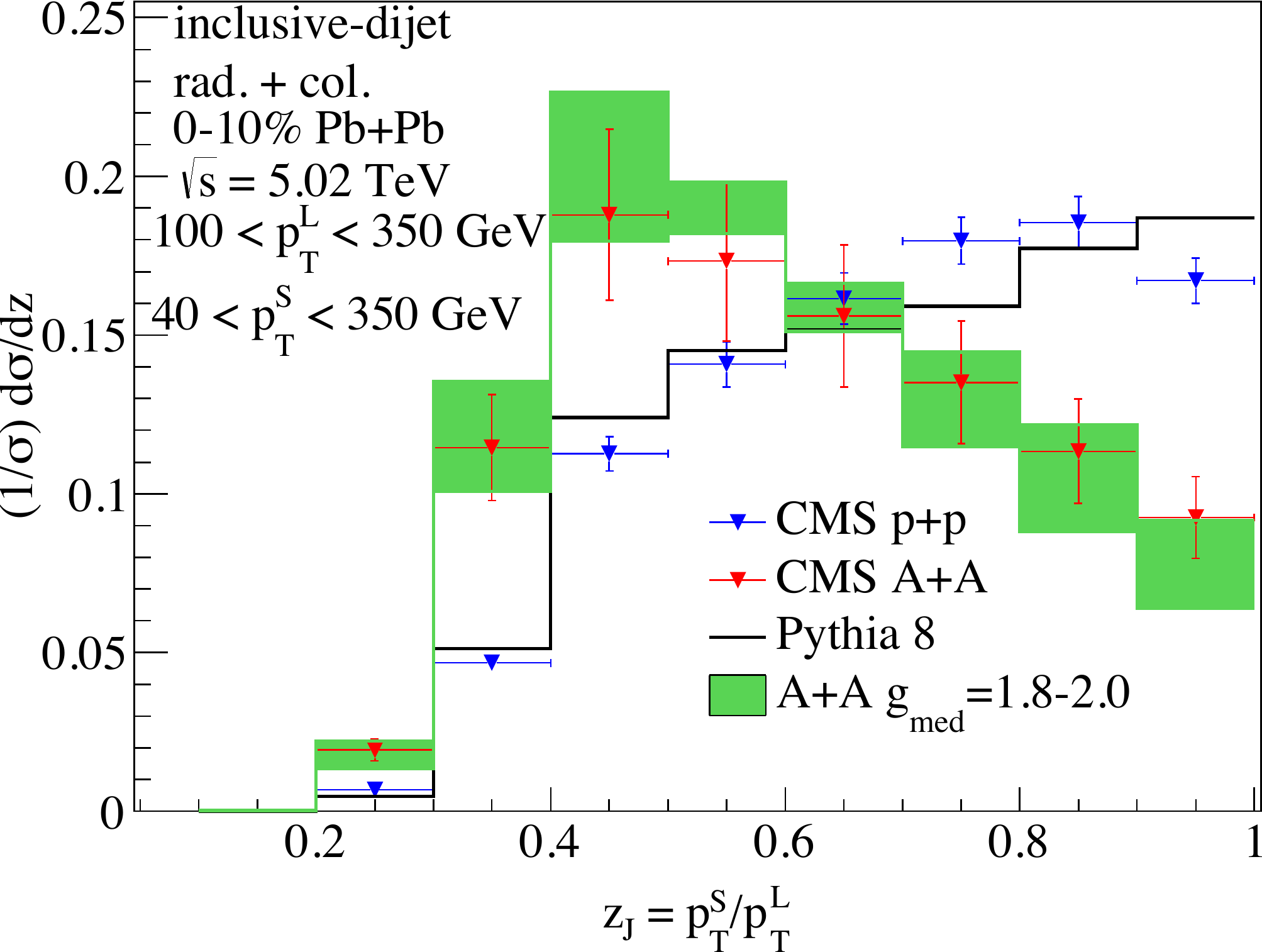}
\hskip 0.2in
\includegraphics[width=3.0in]{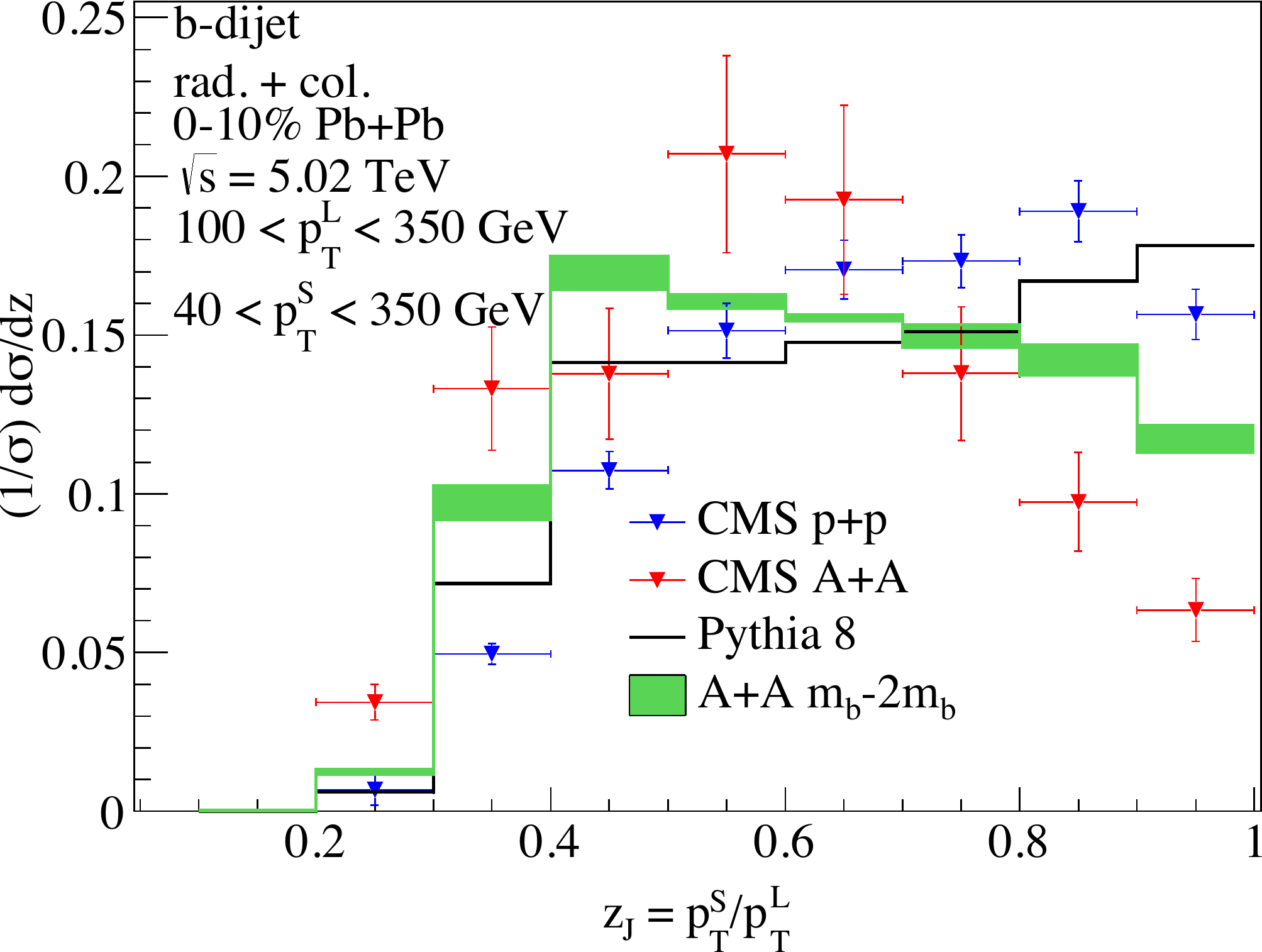}
\caption{The dijet imbalance $z_J$ distributions for inclusive (left) and $b$-tagged (right) dijet production at $\sqrt{s_{NN}} = 5.02$ TeV for CMS at the LHC. The black histogram is the result for p+p collisions, while the colored curves are the results for central ($0-10\%$) Pb+Pb collisions. Left: band corresponds to a range of coupling strengths between the jet and the medium: $g_{\rm med}=1.8-2.0$, respectively. Right: we fix $g_{\rm med}=1.8$, and the band corresponds to a range of masses of the propagating system between $m_b$ and $2m_b$. The experimental data are from CMS collaboration~\cite{Sirunyan:2018jju}.}
\label{fig:CMS_zj}
\eef

To quantitatively compare the medium modification of $b$-tagged and inclusive dijet production, we further plot the ratio of nuclear modification factors for $b$-tagged ($R_{AA}^{bb}$) and inclusive dijet ($R_{AA}^{jj}$) production, $R_{AA}^{bb}/R_{AA}^{jj}$, as a function of dijet invariant mass $m_{12}$ in Fig.~\ref{fig:R_AA_mass_ratio}. The left panel shows the results for central Pb+Pb collisions at LHC energy $\sqrt{s_{NN}} = 5.02$ TeV, while the right panel shows the results for central Au+Au collisions at sPHENIX energy $\sqrt{s_{NN}} = 200$ GeV. For LHC (sPHENIX) energies, we choose $g_{\rm med} = 1.8~(2.0)$. For $b$-tagged dijets, the mass of the propagating system is held fixed at $m_b$. In both kinematic regimes, we see a smaller suppression (thus larger $R_{AA}$) for $b$-tagged dijets compared to inclusive dijets, though the figure also indicates a markedly different effect at low energies than at higher ones. The most pronounced differences occur in the low mass range $m_{12}\sim 20$ GeV accessible by sPHENIX, where such a ratio reaches up to almost a factor of 10, $R_{AA}^{bb}/R_{AA}^{jj}\sim 10$. On the other hand, at LHC energy, one should observe roughly a factor of 2 less suppression for $b$-tagged dijet at relatively low dijet invariant mass $m_{12}$. For large $m_{12}\sim 500$ GeV, the difference diminishes and one should expect to see similar suppressions, $R_{AA}^{bb}/R_{AA}^{jj}\sim 1$.

\bef
\includegraphics[width=3.0in]{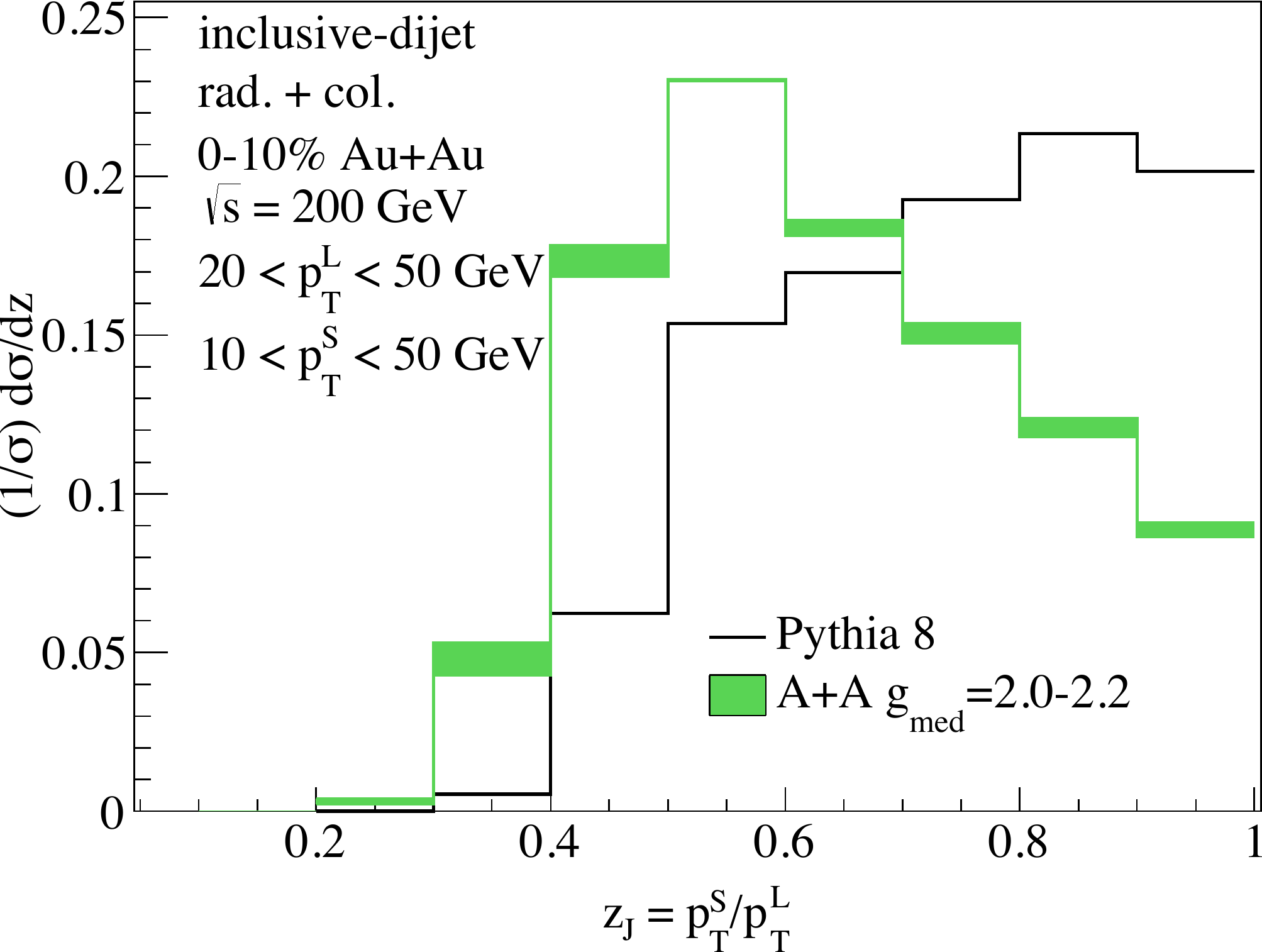}
\hskip 0.2in
\includegraphics[width=3.0in]{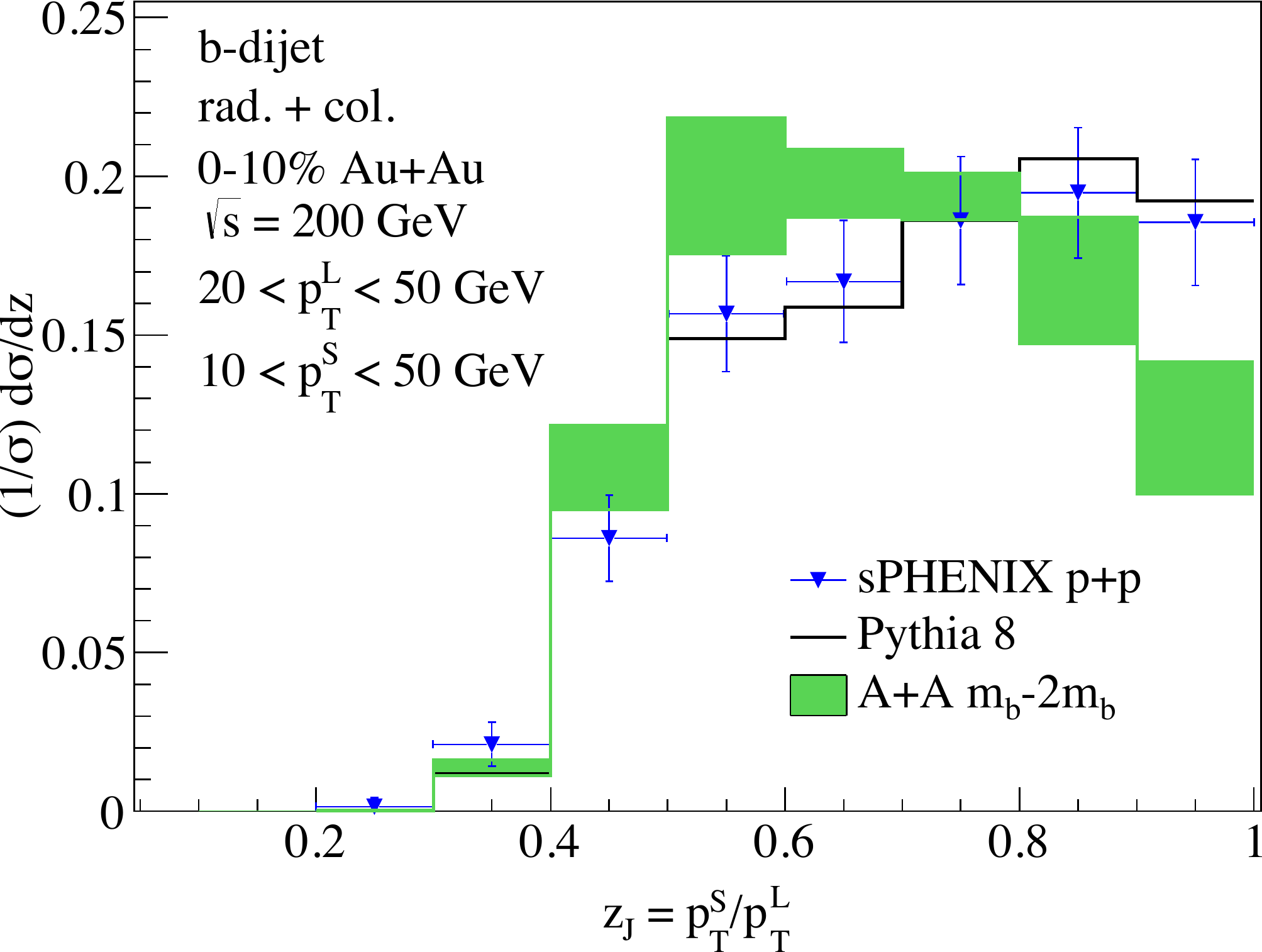}
\caption{The dijet imbalance $z_J$ distributions for inclusive (left) and $b$-tagged (right) dijet production at $\sqrt{s_{NN}} = 200$ GeV for sPHENIX at RHIC. The black histogram is the result for p+p collisions, while the colored curves are the results for central ($0-10\%$) Au+Au collisions. The blue ``data'' points are from preliminary simulations carried out by the sPHENIX collaboration~\cite{sPHENIX}. Left: band corresponds to a range of coupling strengths between the jet and the medium: $g_{\rm med}=2.0-2.2$, respectively. Right: we fix $g_{\rm med}=2.0$, and the band corresponds to a range of masses of the propagating system between $m_b$ and $2m_b$.}
\label{fig:sPHENIX_zj}
\eef

Let us now turn to the conventional observable, the momentum imbalance distributions, $d\sigma/dz_J$. In the absence of in-medium interactions, one expects from perturbative QCD that the transverse momenta of the two jets are balanced, $p_{1T}\approx p_{2T}$. Consequently, $d\sigma/dz_J$ in elementary p+p collisions will be peaked around $z_J\approx 1$. On the other hand, in heavy ion collisions, jet quenching plays an important role and one jet will lose more energy than the other. As a result, one expects to see a downshift of the peak in $z_J$ distribution because of strong in-medium interactions. 

In Fig.~\ref{fig:CMS_zj}, we display the normalized dijet imbalance $z_J$ distributions for inclusive (left) and $b$-tagged (right) dijet production at the LHC energy $\sqrt{s_{NN}}=5.02$ TeV. The black histogram is the result for p+p collisions, while the colored curves are the results for central ($0-10\%$) Pb+Pb collisions. In the left panel, the band corresponds to a range of coupling strengths between the jet and the medium: $g_{\rm med}=1.8-2.0$, respectively. In the right panel, we fix $g_{\rm med}=1.8$, and the band corresponds to a range of masses of the propagating system between $m_b$ and $2m_b$. The experimental data points are from CMS collaboration~\cite{Sirunyan:2018jju}. We clearly see a downshift in the peak of $z_J$ distribution for both inclusive and $b$-tagged dijet production. There is an excellent agreement between our calculations for inclusive dijets and the CMS data. On the other hand, our calculations do not describe very well the CMS data for $b$-tagged dijets. We attribute this to the use of purely LO matrix elements via Pythia 8 and the specific nature of the re-weighting procedure carried out by CMS~\cite{Sirunyan:2018jju}. We do not carry out such a re-weighting procedure in order to maintain consistency with the rest of our simulations. Note that the visual difference between our results in A+A and the experimental data is also largely driven by the p+p baseline. Our calculation with  $g_{\rm med}=2.0$ appears closer to the Pb+Pb data. However, as shown below, the results with $g_{\rm med}=1.8$ already quantitatively capture the downshift of the $z_J$ distribution in heavy ion collisions. This again emphasizes the fact that from the momentum imbalance distributions alone it might be difficult to assess whether a theoretical model correctly represents the physics of jet quenching.
Fig.~\ref{fig:sPHENIX_zj} contains the dijet imbalance $z_J$ distributions for inclusive (left) and $b$-tagged (right) dijet production with sPHENIX kinematics at $\sqrt{s_{NN}} = 200$ GeV. Our results for $b$-tagged dijets in p+p collisions are consistent with the preliminary simulations carried out by the sPHENIX collaboration~\cite{sPHENIX} (denoted as the blue ``data'' points). Our calculations show that a larger shift in $z_J$ should be observed for inclusive dijets compared with $b$-tagged dijets. 

\begin{table}
\caption{Theoretical results for the difference of the average dijet imbalance $z_J$ between p+p and Pb+Pb collisions at $0-10\%$ centrality (CMS) and Au+Au collisions at $0-10\%$ centrality (sPHENIX). Results for CMS may be compared to the experimentally measured values. For both kinematics, we observe a larger shift in imbalance for light flavor dijets than for their heavy counterparts. Both inclusive and $b$-tagged ranges correspond to the values obtained by varying the coupling to the medium. For CMS: $g_{\rm med}=1.8-2.0$. For sPHENIX: $g_{\rm med}=2.0-2.2$, where the mass of the propagating system is held fixed at $m_b$.}
\label{table:downshift}
\begin{center}
\begin{tabular}{c | c | c | c | c}
\multicolumn{5}{c}{} \\
\hline
\hline
Kinematics & dijet flavor & $\langle z_{J} \rangle _{\rm pp}$ & $\langle z_{J} \rangle _{\rm AA}$ & $\Delta \langle z_{J} \rangle$ \\
\hline

CMS~\cite{Sirunyan:2018jju} & b-tagged & $0.661 \pm 0.003 $ & $0.601 \pm 0.023$ & $0.060 \pm 0.025$ \\
Experiment & inclusive & $0.669 \pm 0.002$ & $0.617 \pm 0.027$ & $0.052 \pm 0.024$ \\
& & & \\

LHC & b-tagged & 0.685 & $0.626 \pm 0.013$ & $0.059 \pm 0.013$ \\
theory& inclusive & 0.701 & $0.605 \pm 0.022$ & $0.096 \pm 0.022$     \\
& & &  \\

sPHENIX & b-tagged & 0.730 & $0.665 \pm 0.012$ & $0.065 \pm 0.012$ \\
theory & inclusive & 0.743 & $0.643 \pm 0.005$ & $0.100 \pm 0.005$    \\
\hline
\hline
\end{tabular}
\end{center}
\end{table}

To further quantify the downshift of the $z_J$ distribution, we define the mean value of $z_J$,
\bea
\langle z_J \rangle = \left.\left(\int dz_J \, z_J\frac{d\sigma}{dz_J}\right)\right/\left(\int dz_J \frac{d\sigma}{dz_J}\right).
\eea
We further define the difference for $\langle z_J \rangle$ in p+p and A+A collisions as
\bea
\Delta \langle z_J\rangle = \langle z_J\rangle_{\rm pp} - \langle z_J\rangle_{\rm AA},
\eea
and the positive values of $\Delta \langle z_J\rangle$ represents downshifts of the $z_J$ distribution in A+A collisions in comparison with that of the p+p collisions. In Table~\ref{table:downshift}, we list our theoretical calculations for $\langle z_J\rangle_{\rm pp}$, $\langle z_J\rangle_{\rm AA}$, and $\Delta \langle z_J\rangle$ for both inclusive and $b$-tagged dijet production. The values labelled as ``LHC theory'' are our theoretical calculations for Pb+Pb collisions at $0-10\%$ centrality at $\sqrt{s_{NN}} = 5.02$ TeV, and can be compared with the CMS experimental data. For inclusive dijets, we perform the calculations for the coupling between the jet and the medium $g_{\rm med}=1.8-2.0$, which explains the uncertainties in our theoretical values. For $b$-tagged dijets, we vary such a coupling in the same range while the mass of the propagating system is held fixed at $m_b$. 
We find that in general the downshift $\Delta \langle z_J\rangle$ is slightly larger for inclusive dijet production than that for $b$-tagged dijets, though the uncertainties are still large. Nevertheless, within the theoretical and experimental uncertainties, our theoretical calculations for all these observables $\langle z_J\rangle_{\rm pp}$, $\langle z_J\rangle_{\rm AA}$, and $\Delta \langle z_J\rangle$, agree well with the CMS experimental data. Finally, we also perform calculations for central Au+Au collisions for sPHENIX kinematics at $\sqrt{s_{NN}} = 200$ GeV in Table~\ref{table:downshift}, which are labelled as ``sPHENIX theory.'' We expect such measurements will become available once the sPHENIX experiment starts running in the future. 

\section{Conclusions}

In this paper we present detailed theoretical predictions for inclusive and $b$-tagged dijet production and modification in heavy ion collisions at RHIC and the LHC. We propose a new observable, the modification of dijet invariant mass, as a novel diagnostic of the quark-gluon plasma (QGP) created in ultra-relativistic heavy ion collisions. Our comprehensive studies conclusively demonstrate that this observable exhibits enhanced sensitivity to the strength of jet-medium interactions, the transport properties of nuclear matter, and to the mass effects on in-medium parton showers. Complete characterization of the quenching of multi-jet events is given by a multi-dimensional nuclear modification ratio, as we also show in this work. The statistics necessary to perform such measurements at present, however, makes them impractical even for the case of two jets. By integrating out one of those dimensions, which is usually accomplished through an auxiliary variable such as the dijet momentum asymmetry or dijet momentum imbalance, the statistics can be greatly improved and experimental results may be obtained. Such traditional momentum imbalance measurements emphasize potentially small differences in the quenching of leading and subleading jets. Hence, the shift in the mean value of the momentum imbalance variable $z_J$ is only on the order of $7-15$\%. Differences between the dijet asymmetries of inclusive and $b$-tagged dijets have been difficult to identify. In contrast, the dijet mass modification combines the suppression of the individual jets and enhances the observable jet quenching effect by up to an order of magnitude.

To obtain reliable predictions, we use radiative and collisional parton energy losses evaluated in a realistic hydrodynamic background. Furthermore, we utilize couplings between the jet and the medium that have successfully predicted or described the nuclear modification of a number of observables related to the reconstructed jets in heavy ion collisions - inclusive light and heavy flavor jet suppression, jet substructure, and vector-boson-tagged jets. We find that the theoretical model gives an excellent description of the most recent measurements of the inclusive dijet momentum imbalance distributions and their modification in heavy ion collisions at the LHC. For the $b$-tagged dijets the description is not nearly as good, though the theoretical model qualitatively and even quantitatively captures the shift of the momentum imbalance in Pb+Pb relative to p+p collisions. The deviation in shape can likely be attributed to the lowest order Pythia 8 baseline simulation.

For the main result of this paper, the dijet mass distribution modification, we find that the suppression at $m_{12} \sim 100$~GeV is around a factor of 10. In contrast, the suppression of single inclusive jets is only around a factor of 2. We note that as the dijet mass grows, the suppression is reduced and at $m_{12} \sim 500$~GeV, it is about a factor of 2. When the nuclear modification due to final-state interactions diminishes, initial-state cold nuclear matter effects may play a more important role, see for example the effect of cold nuclear matter energy loss~\cite{Kang:2015mta}. In the regime of small parton momentum fraction $x$, the nonlinear gluon saturation effect could also be important~\cite{Gelis:2010nm}, although we are interested in the high transverse momentum region which typically has moderate to large $x$. We defer such studies for the future after the first experimental measurements of dijet mass modification appear. In the mass region studied for the LHC, the modification of the $b$-tagged dijet mass distribution can be twice as small as that of inclusive dijets, though this difference disappears at larger masses.    

Since sPHENIX at RHIC is expected to become available in the near future, we also perform calculations of dijet mass distributions and momentum imbalance distributions. We find that jet quenching effects on the dijet mass distribution can be significantly amplified in the kinematic range accessible by the future sPHENIX experiment, because of steeply falling spectra. In the mass region $m_{12} = 20 - 100$~GeV, the QGP-induced suppression is a factor of 10 or larger for inclusive dijet production. A 10\% change in the strong coupling constant $g$ that describes the jet-medium interactions can lead to a factor of 2 larger suppression. On the other hand, the suppression for $b$-tagged dijets shows a different behavior, which can be traced back to the heavy quark mass effects. In other words, at sPHENIX kinematics, there is an enhanced sensitivity to heavy quark mass effects, and we find that in the smaller dijet mass range the suppression for $b$-tagged dijets can be an order of magnitude smaller. 

To conclude, upcoming runs at RHIC and the LHC present compelling opportunities for experiments to explore novel jet quenching observables. The modification of light and heavy flavor dijet mass distributions will be a promising avenue of exploration in this direction. 

\section*{Acknowledgments}
We thank Jin Huang and Hongxi Xing for fruitful discussions. Z.~Kang and J.~Reiten thank the theoretical division at Los Alamos National Laboratory for its hospitality and support where part of this work was performed. Z.~Kang and J.~Reiten are supported by the National Science Foundation under Grant No.~PHY-1720486 and by a UCLA faculty research grant. I.~Vitev is partly supported by U.S. Department of Energy under Contract No.~DE-AC52-06NA25396, the DOE Early Career Program, and the LDRD program at LANL. B.~Yoon is supported by the LANL LDRD program.

\bibliographystyle{h-physrev5}   
\bibliography{bibliography}

\end{document}